\documentclass[journal]{IEEEtran}
\usepackage{graphics} % for pdf, bitmapped graphics files
\usepackage{graphicx} % for pdf, bitmapped graphics files
\usepackage{epsfig} % for postscript graphics files
\usepackage{mathptmx} % assumes new font selection scheme installed
\usepackage{times} % assumes new font selection scheme installed
\usepackage{amsmath} % assumes amsmath package installed
\usepackage{amssymb}  % assumes amsmath package installed
\usepackage{cite}
\usepackage{lipsum}
\usepackage{adjustbox}
\usepackage{multirow}
\usepackage{hyperref}
\usepackage{textcomp}
\usepackage{mathtools}
\usepackage{algorithmicx}
\usepackage[ruled]{algorithm}
\usepackage{algpseudocode}
\usepackage{xcolor}
\algrenewcommand\alglinenumber[1]{\tiny #1:}
\usepackage{caption} 

\begin{document}
	\title{Deep MR Brain Image Super-Resolution Using Spatio-Structural Priors}
	
	\author{Venkateswararao Cherukuri$^{1,2}$, Tiantong Guo$^{1}$, Steven J. Schiff $^{2, 3}$, Vishal Monga$^{1}$% stops a space
		\thanks{*This work is supported by NIH Grant number 7R01HD085853.}% <-this % stops a space
		%\thanks{$^{\dag}$Contributed equally.}% <-this % stops a space
		\thanks{$^{1}$Dept. of Electrical Engineering, $^{2}$Center for Neural Engineering, $^{3}$Dept. Neurosurgery, Engineering Science and Mechanics, and Physics, The Pennsylvania State University, University Park, USA.
		}%
		%\thanks{$^{4}$CURE Children\textquotesingle s Hospital of Uganda, Mbale, Uganda, $^{5}$Department of Neurosurgery, Boston Children\textquotesingle s Hospital and Department of
		%Global Health and Social Medicine, Harvard Medical School, Boston, Massachusetts, $^{6}$Division of Neurosurgery, Hospital for Sick Children, University of Toronto, Toronto, ON, Canada
		
	}% <-this % stops a space
	%\thanks{M. Shell was with the Department
	%of Electrical and Computer Engineering, Georgia Institute of Technology, Atlanta,

	%\markboth{IEEE TRANSACTIONS ON BIOMEDICAL ENGINEERING}%
	
	\maketitle
	
	\begin{abstract}
		High resolution Magnetic Resonance (MR) images are desired for accurate diagnostics. In practice, image resolution is restricted
		by factors like hardware and processing constraints. 
		Recently, deep learning methods have been shown
		to produce compelling state-of-the-art results for image enhancement/super-resolution. Paying particular attention to desired hi-resolution MR image structure, we propose a new regularized network that exploits image priors, namely a low-rank structure and a sharpness prior to enhance deep MR image super-resolution (SR). Our contributions are then incorporating these priors in an analytically tractable fashion  \color{black} as well as towards a novel prior guided network architecture \color{black} that accomplishes the super-resolution
		task. This is particularly challenging for the low rank prior since the rank is not a differentiable function of the image matrix
		(and hence the network parameters), an issue we address by pursuing differentiable approximations of the rank. Sharpness
		is emphasized by the variance of the Laplacian which we show can be implemented by a fixed feedback layer at the output
		of the network. As a key extension, we modify the fixed feedback (Laplacian) layer by learning a new set of training data driven filters that are optimized for enhanced sharpness. Experiments performed on publicly available MR brain image databases and comparisons against existing state-of-the-art methods show that the proposed prior guided network offers significant practical gains in terms of improved SNR/image quality measures. Because our priors are on output images, the proposed method is versatile and can be combined with a wide variety of existing network architectures to further enhance their performance.
	\end{abstract}
	
	
	\IEEEpeerreviewmaketitle
	
	\section{Introduction}
	\label{sec:intro}
	High Resolution (HR) MR images provide rich structural information about bodily organs which is critical in analyzing any given medical condition. Often, the quality of these images is restricted by factors like imaging hardware, sensor noise, budget, and time constraints. In such scenarios, the spatial resolution of these images can be enhanced by a well-designed mathematical algorithm. Simple and fast interpolation methods like bilinear and bicubic \cite{lehmann1999survey} have been widely used for increasing the size of low-resolution (LR) medical images. In many cases, these methods are known to introduce blurring, blocking artifacts, ringing and are thus unable to recover sharp details of an image. To alleviate this problem, an alternative approach known as super-resolution (SR) was introduced in \cite{tsai1984multiframe}. Current literature on SR can be classified into two categories: multi-image SR and single-image SR.\\
	In multi-image SR \cite{tsai1984multiframe, farsiu2004fast}, an HR image is generated by exploiting the information from multiple LR images which are acquired from the same scene with a slightly shifted field of view. However, these methods are likely to fail if an adequate amount of LR images from the same scene are not available. As an alternative approach, single image SR was introduced wherein multiple LR images from the same scene are not required to obtain an HR image. In this approach, a mapping between LR and HR images is learned by constructing examples from a given database \cite{trinh2014novel, freeman2002example, chang2004super, yang2010image, bahrami2016reconstruction, wen2018power}. \\
	%Single image SR is achieved by using a single LR image by estimating the degradation model using prior information on medical images \cite{shi2015lrtv, zhao2016single, rousseau2010non, manjon2010non} or by using multiple LR images which are not necessarily from the same scene to learn a mapping from LR to HR images which are also known as example based learning methods \cite{trinh2014novel, freeman2002example, chang2004super, yang2010image}. \\
	Recently, deep learning methods have been shown to produce compelling state-of-the-art results \cite{lim2017enhanced,dong2016image,kim2016accurate,liu2016robust,guo2017deep,wang2015deep,dong2016accelerating,timofte2017ntire,kim2016deeply,tiantong2018ortho,timofte2018ntire, TIP2019, zhang2018residual, zhang2018image} for single image SR.  Invariably though, the training requirement of deep networks, i.e. the number of example LR and HR image (or patch) pairs, is quite significant. In some medical diagnosis problems, generous LR and HR pairs is not a problem but there are compelling real-world problems such as enhancing 3T MR to 7T MR images \cite{bahrami2016reconstruction,bahrami2016convolutional}, where the paucity of training has been recognized.  There has been encouraging recent application of deep networks for MR image SR \cite{chen2018brain, shi2018super, yang2016super, srinivasan2017super, pham2017brain} but the methods remain training intensive. An outstanding open challenge for deep MR image super-resolution is the development of methods that exhibit a graceful degradation with respect to (w.r.t.) the number of training LR and HR image pairs. \\
	\begin{figure}[t]
		\centering
		\begin{center}
			\includegraphics[scale=.3]{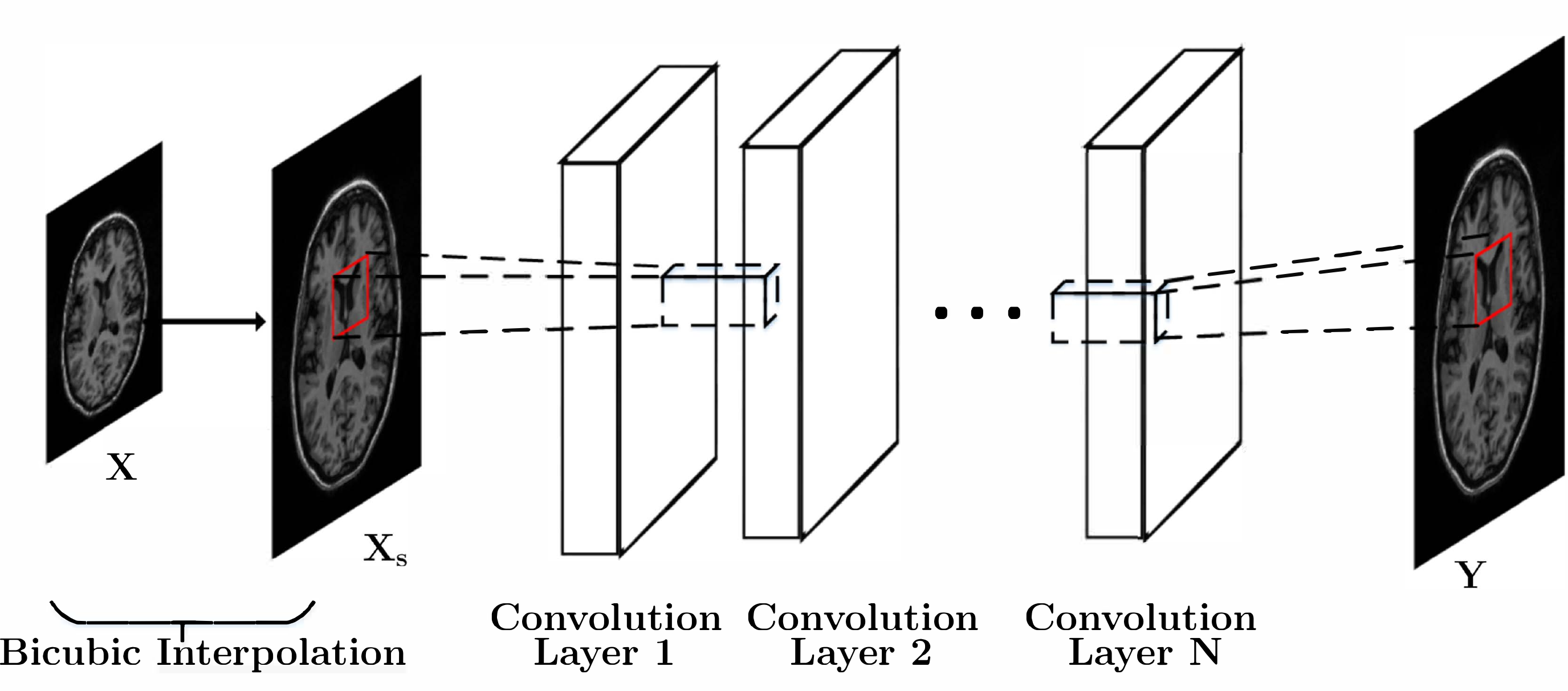}
		\end{center}
		\caption{\small{Super-Resolution CNN (SRCNN).}}
		\label{fig:SRCNN}
	\end{figure}
	Our approach to improve deep MR image superresolution for all training regimes is via the exploitation of suitable structural prior information pertinent to MR images. In \cite{shi2015lrtv}, a model based SR approach is presented that uses low-rank (approximated by nuclear norm) and total variation regularizers. %The authors in \cite{shi2015lrtv} validate that MR images from various parts of the body can be reconstructed with a peak signal-to-noise ratio (PSNR) of close to $50$ db by retaining only half of the singular values obtained by a singular value decomposition (SVD) of the image matrix. 
	Despite the promise shown by a low rank prior, incorporating a low rank constraint or even its nuclear norm relaxation in a deep network for SR presents a stiff analytical challenge since neither is a differentiable function of the image matrix (and hence the network parameters). Our contribution includes incorporating a suitable approximation to the rank, which is smooth, differentiable and amenable for learning in a deep CNN framework. Additionally, recognizing the need for well formed sharp edges in diagnosis, we propose a sharpness prior realized via a variance of the Laplacian measure which adds to the network structure at the output as a fixed feedback layer.  As we bring prior information into a deep network, we call our method Deep Network with Spatio-Structural Priors (DNSP). As a key extension, we modify the fixed feedback (Laplacian) layer by learning a new set of training data driven filters that are optimized for enhanced sharpness.
	
	Another class of approaches for single image SR called self-image SR \cite{shocher2018zero,huang2015single} has been developed recently. Self image SR has been adapted to enhance resolution of MR images \cite{jog2016self,zhao2018self}. These approaches exploit the fact that MR images are inherently anisotropic to learn a regression between LR and HR images. Thus, they generate new additional images, each of which is LR along a certain
	direction, but is HR in the plane normal to it. Further, these approaches improve the resolution across z axis assuming the images in axial plane are HR. But, in our approach, we focus on improving the resolution of images in axial plane by exploiting suitable structural information of MR images thereby differentiating from the aforementioned methods.\\ 
	\textbf{Contributions:} While most of the existing deep learning methods for MR image SR focus on learning an end to end relationship between LR and HR image, our goal is to enrich this deep learning framework by bringing in suitable structural information via informative priors and incorporating them in an analytically tractable fashion. Specifically, our key contributions are as follows:
	\begin{itemize}
		
		\item \textbf{Novel Prior-Guided Network Structure:}  \color{black} We propose a new network structure that consists of two components: 1) A regression network that maps LR to HR images 2.) a {\em prior information network} component that guides the learning of the regression network during the training phase. Note only, the regression network is used in the test (inference) stage to map LR to HR images. \color{black}
		
		\item \textbf{Incorporating Spatio-Structural Priors:} We impose a low rank constraint on the output of the deep network. Evidence for brain MR images being low-rank has been provided recently \cite{shi2015lrtv}. However, incorporating a low rank constraint or even its nuclear norm relaxation into a deep learning framework is not straightforward as neither of the functions are differentiable. We provide a solution to integrate low rank constraint into a deep network by approximating the rank function with a smooth and differentiable function. We further incorporate a spatially based sharpness prior defined as the variance of Laplacian computed on the network output (image). Laplacian can be implemented via a linear convolution with a filter (fixed) and subsequently, the variance is computed to yield a regularization term.
		\item \textbf{Data Adaptive Filters to Enhance Sharpness:} We further extend the aforementioned two contributions by learning a series of filters that are aimed at enhancing the sharpness of the output image. We develop new data adaptive regularizers which ensure that the learned sharpness filters are physically meaningful.
		\item \textbf{Novel Regularized Loss Function:} Analytically, to integrate the proposed priors, we introduce three new regularization terms in the loss function along with the standard reconstruction loss term. The first regularization term poses a low rank prior, the second is a sharpness prior, while the third constrains the filters that replace the Laplacian and are aimed at enhancing sharpness. Further, back-propagation equations for optimizing  network parameters w.r.t the regularized loss function are derived in a form that is implementation friendly.  
		\item \textbf{Experimental validation and reproducibility:} Experimental validation of our method is carried out on two publicly available data bases: 1.) Brainweb (BW)\footnote{\url{http://brainweb.bic.mni.mcgill.ca/brainweb/}} and 2.) Alzheimer's Disease Neuroimaging Initiative (ADNI)\footnote{\url{http://adni.loni.usc.edu/}}. We compare DNSP against several state of the art methods that are used for MR image SR. We also provide the entire code of our experiments for the purpose of reproducibility at  \url{https://scholarsphere.psu.edu/concern/generic_works/9s4655g25h}.        
	\end{itemize}
	A preliminary version of this work was presented as a 4 page conference paper at 2018 IEEE International Conference on Image Processing  \cite{cherukuri2018deep}\footnote{Our extensions from conference (4 pages) to Journal (13 pages) are consistent with the IEEE signal processing society guidelines: \url{https://signalprocessingsociety.org/publications-resources/information-authors}}. This present draft involves both substantial conceptual and experimental extensions including: 1.) We evolve the fixed Laplacian layer into a learnable one that computes an enhanced sharpness measure, 2.) more detailed analytical development is presented including back-propogation derivations under the new regularizers, 3) we integrate the aforementioned priors with the most competitive deep network architectures that are used for image SR thereby demonstrating the versatility of our method, and 4) we significantly expand experiments by comparing against many new state-of-the-art methods. Results are also presented for several variants of DNSP and in many new scenarios (training and test selection). \\
	The rest of the paper is organized as follows. The proposed prior guided deep network for MR image SR is explained in Section \ref{sec:DNSP}. Extensions of DNSP to include learnable sharpness filters is subsequently presented in Section \ref{sec:LALF}. Detailed experimental validation against the state of the art methods is reported in Section \ref{sec:Experiments}. Finally, our work is summarized and concluded in Section \ref{sec:Conclusions}.\\
	
	\section{Spatio-Structural Priors for Deep MR image super-resolution}
	\label{sec:DNSP}
	%\vspace{-.3cm}
	We first introduce the notation that is followed through rest of the paper and then give a brief introduction for deep networks for image SR before describing our DNSP method.   
	\subsection{Notation}
	%\vspace{-.3cm}
	Let $X \in \mathbb{R}^{M\times N}$ represent the LR image where $M$ and $N$ are the width and height of the image respectively. Let $Y\in \mathbb{R}^{sM\times sN}$ be the output HR image and $s$ is the desired scale to which $X$ needs to be upscaled and $Y_{g} \in \mathbb{R}^{sM\times sN}$ is the ground truth HR image for $X$. Let $W_{k}^{l} \in \mathbb{R}^{m\times n\times d}$ be the $k^{th}$ convolutional filter in layer $l$ where $m$, $n$ and $d$ represent the width, height and depth of the filter respectively. Similarly, let $b_{k}^{l} \in \mathbb{R}$ be the $k^{th}$ bias coefficient of layer $l$. The objective of the network is to learn $W_{k}^{l}$ and $b_{k}^{l}$ so that the output of the network $Y$ is a close representation of the ground truth $Y_{g}$. So, let $\Theta = \{W_{k}^{l}, b_{k}^{l}\} \forall l,k$. To make the size of input and output of the network the same, we first upscale $X$ by a factor of $s$ using bicubic interpolation and use this upscaled $X_{s}\in \mathbb{R}^{sM\times sN}$ as input to the network. Finally, let the mapping function of the network be represented by $F$ where $F(X_{s}, \Theta) = Y$.
	
	\subsection{Deep CNNs For SR}
	%\vspace{-.3cm}
	Deep learning methods are a class of machine learning methods which are inspired by biological neural networks. In general, a cascade of many nonlinear processing units are used to learn features to represent data effectively for a  given task. In particular, a deep CNN for image SR usually consists of two or more convolutional layers (each layer essentially is a combination of filters followed by an activation function) which are used to learn an end-to-end  mapping between sample HR and LR image pairs. For example, Fig. \ref{fig:SRCNN} illustrates the SRCNN network \cite{dong2016image, yang2016super} for super-resolution which is known to be the most widely used deep SR network. Following these footsteps, many new architectures have been designed for image SR which showed considerable gains in terms of performance \cite{lim2017enhanced,kim2016accurate,guo2017deep,wang2015deep,dong2016accelerating,timofte2017ntire,kim2016deeply, TIP2019, zhang2018residual, zhang2018image}.  Each convolutional layer in the network consists of several learnable filters, which are convolved with the output from the previous layer. For a given layer, outputs obtained by convoluting with each filter are combined to form a data cube which is passed through a nonlinear activation function and then forwarded as an input to next layer \cite{lecun2015deep}. Most commonly used activation function in deep networks is the Rectified linear unit (Relu) \cite{glorot2011deep}. The input to the first layer is the image obtained after bicubic interpolation and the output of the last layer is the expected HR image. The filters are learned to minimize the loss function given by:
	\begin{equation}\label{eq1}
	E(\Theta) = \frac{1}{2}\|Y_{g} - F(X_{s}, \Theta)\|_{F}^{2}
	\end{equation}
	where $\parallel\cdot\|_{F}$ represents the Frobenius norm.
	
	\subsection{Deep Network with Spatio-Structural Priors (DNSP)}
	\label{sec:DNSP_fix}
	As discussed in Section \ref{sec:intro}, we integrate two priors into the learning of the CNN. Note that both the priors are to be applied on $Y$ as it represents the desired output HR image. The two priors are as follows: \\
	
	\begin{figure}[t]
		\begin{center}
			\includegraphics[width=\linewidth]{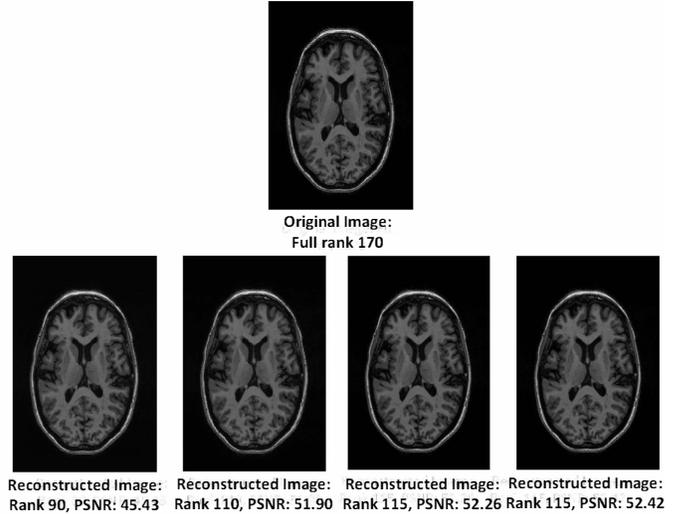}
		\end{center}
		%	\vspace{-.5cm}
		\caption{ \color{black}\small{An example that demonstrates that MR brain images are naturally rank deficient. The low rank images are obtained by zeroing out the smallest singular values (from the SVD). This  example reveals that the image has an effective rank in the range 115-120.\color{black}}}
		\label{fig:rankDemo}
	\end{figure}
	
	\textbf{Low Rank Prior:} It has been demonstrated recently \cite{shi2015lrtv, 8332964, wen2018power, tang2017groupwise} that MR images are naturally rank deficient. For example, Figure \ref{fig:rankDemo} shows several low rank images of an MR image reconstructed from partial singular value decomposition (SVD) approximation. We can observe that the recovered image with a rank of 90, which is approximately half of full rank (170) of the image matrix; still exhibits a Peak Signal to Noise Ratio (PSNR) of about 45dB. Further, the reconstruction is visually indistinguishable from the original image.  \color{black} We wish to emphasize that an image being low-rank implies that the effective rank of the matrix is low. For example, it can be observed from Figure \ref{fig:rankDemo} that the change in PSNR value in the range of 110-120 rank is relatively negligible compared to that of the PSNR change in the range of 90-110 rank. Hence the effective rank of this particular image can be argued to be in between 115 and 120 which is much smaller than the full rank of 170. Rank of an image captures the global structure of a given image. An effective low-rank implies that the image adheres to some structural properties like near symmetry which can be observed in brain images. Hence, a low-rank constraint is effective in recovering the global structure of a given brain image.     
	
	\color{black}
	However, the rank of a matrix is a non-differentiable function w.r.t. its input and therefore cannot be used as regularizer in a CNN. Most of the optimization problems with a low-rank constraint are solved by minimizing the nuclear norm of the matrix which is a convex relaxation of the low-rank constraint. However, this relaxation also cannot be used in a CNN as the nuclear norm is also a non-differentiable function. To address this, we pursue smooth and differentiable approximations of the rank. In particular, in recent work \cite{malek2014recovery} an estimate of the number of singular values of a matrix that are zero was proposed as:
	\begin{equation}\label{eq2}
	G_{\delta}(Y) = h_{\delta}(\pmb{\sigma}(Y))
	\end{equation}
	where $h_{\delta}(\pmb{\sigma}(Y)) = \sum_{i=1}^{R}g_{\delta}(\sigma_{i}(Y))$, $\sigma_{i}(Y)$ represents the $i^{th}$ singular value of $Y$ and
	\begin{equation}\label{eq3}
	g_{\delta}(x) = \exp\Bigg(-\frac{x^{2}}{2\delta^{2}}\Bigg)
	\end{equation}
	where $\delta$ is a tunable parameter that affects the measure of approximation error in finding the rank. Intuitively, for small $\delta$, $G_{\delta}(Y)$ gives the number of singular values of $Y$ which are zero. Therefore, $rank(Y)\approx R - G_{\delta}(Y)$. Let $R_{\delta}(Y) = R - G_{\delta}(Y)$, where $R = \min(sM, sN)$. Now, the function $R_{\delta}(Y)$ is differentiable and its gradient w.r.t. $Y$ is given by:
	\begin{equation}\label{eq4}
	-U\mbox{diag}\Bigg(-\frac{\sigma_{1}}{\delta^{2}}e^{-\sigma_{1}^{2}/2\delta^{2}},\ldots,-\frac{\sigma_{R}}{\delta^{2}}e^{-\sigma_{R}^{2}/2\delta^{2}}\Bigg)Z^{T}
	\end{equation}
	where SVD of $Y = U\mbox{diag}(\sigma_{1}, \ldots, \sigma_{R})Z^{T}$. \\ %$R_{\delta}(Y)$ is integrated to CNN as a regularizer to achieve low rank constraint. \\
	\begin{figure}
		\begin{center}
			\includegraphics[width=\linewidth]{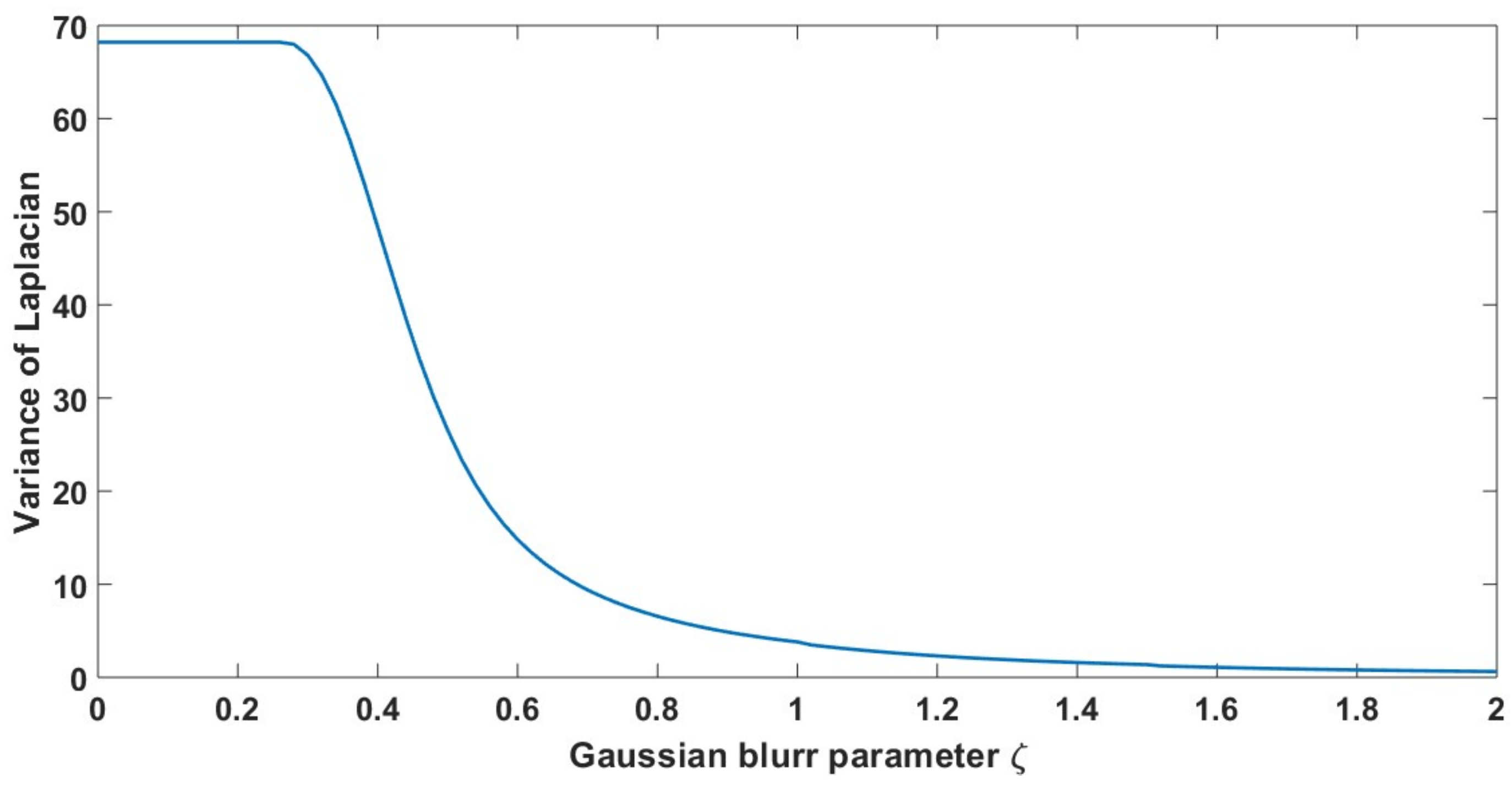}
		\end{center}
		\caption{\small{Variance of the Laplacian vs increasing the blur parameter.}}
		\label{fig:sharpDemo}
	\end{figure}
	\begin{figure*}[t]
		\begin{center}\centering
			\includegraphics[width=0.9\textwidth]{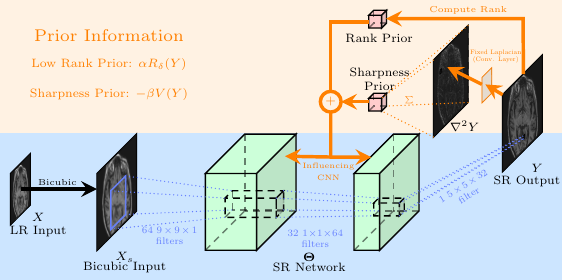}
		\end{center}
		%\vspace{-.5cm}
		\caption{\small{Deep Network with Structural Priors (DNSP) for MR image super-resolution. Note the prior processing (shown in orange) is used {\em only} in the learning of the network. For a given test LR input image $X$, the learned SR network is used to generate the output HR image $Y$. }}
		\label{fig:Net}
	\end{figure*}
	\textbf{Sharpness Prior:} HR images look much sharper compared to LR images. The main reason can be attributed to blurriness of the LR images. The pursuit of quantifying sharpness begins by computing the Laplacian ($\nabla^{2}Y$) of the image \cite{forsyth2011computer}. The laplacian of a smooth/blurred image is more uniform compared to the laplacian of a sharp image. The variance of the Laplacian is hence an indicator of sharpness. As shown in Figure \ref{fig:sharpDemo}, an MR brain image is degraded by a gaussian filter with different blur parameters $\zeta$ and plotted against the variance of laplacian. It can be observed that the variance of laplacian decreases as the blur parameter increases. Therefore, we propose to use $V(Y) = var(\nabla^{2}Y)$ as a regularizer to encourage the CNN to yield sharper HR images. $V(Y)$ is a quadratic function in Y and therefore a differentiable function which can be easily integrated into the CNN learning. Note that the laplacian of an image can be implemented by well-known linear filters \cite{forsyth2011computer}, which are also easily integrated into the CNN via a filtering layer at the output as shown in Fig. \ref{fig:Net}. \\
	
	\noindent \textbf{Remark}:  \color{black} The two priors are chosen carefully so that they perform a complementary job to each other. For example, the low-rank constraint captures the global structure of the brain image and the sharpness prior aids in recovering the finer local structure thereby complementing the low-rank prior. \color{black} \\
	
	\noindent \textbf{Network Structure:} A key advantage of using priors on the output HR image is that they can be incorporated into any network architecture. In Figure \ref{fig:Net}, we show an example where the aforementioned two priors are incorporated into the basic SRCNN \cite{dong2016image} framework. In Section \ref{sec:Experiments}, we demonstrate the versatility of our approach by incorporating the priors with more advanced networks. %Note that our priors can be integrated into any other deep SR network as well. There are 3 layers in DNSP: the first layer has $64$ $9\times 9\times 1$ filters, the second layer has $32$ $1\times 1\times 64$ filters while the final layer has one $5\times 5\times 32$ filter. 
	%The output of each layer except that of the final layer is fed into ReLU to generate a nonlinear activation map \cite{nair2010rectified}. 
	As observed from the Figure \ref{fig:Net}, to obtain the variance of laplacian, we use a $3\times 3$ filter $L = [[0 \mbox{ }- 1 \mbox{ }0]^{T} [-1 \mbox{   4 } \mbox{ }-1]^{T} [0\mbox{ } - 1\mbox{ } 0]^{T}]$ after the final layer to compute the Laplacian and subsequently find the variance of Laplacian. The loss function of DNSP to be minimized is given by:
	\begin{equation}\label{eq5}
	E(\Theta) = \underbrace{\frac{1}{2}\|Y_{g} - F(X_{s}, \Theta)\|_{F}^{2}}_{MSE} + \underbrace{\alpha R_{\delta}(Y)}_{Low Rank} - \underbrace{\beta V(Y)}_{Sharpness Prior}
	\end{equation}
	where, $Y = F(X_{s}, \Theta)$, $\alpha$ and $\beta$ are positive regularization parameters%\footnote{We chose $\alpha = .1$ and $\beta = 5\times 10^{-5}$ by cross validation.}
	, note that negative sign before $V(Y)$ is to increase the variance of Laplacian. Note that the loss function in Eq (\ref{eq1}) is a special case of Eq. (\ref{eq5}). We learn $\Theta$ by minimizing $E(\Theta)$ using a stochastic gradient descent method \cite{lecun1998gradient, werbos1994roots}. In particular, weights are updated by the following equation:
	\begin{equation}\label{eqUpdate}
	\Theta^{t+1} = \Theta^{t} - \eta\frac{\partial E}{\partial\Theta^{t}}
	\end{equation}
	where, $t$ represents the iteration number, $\eta$ represents the learning rate, and $\Theta^{t}$ represents the values of weights at previous iteration. As $\Theta = \{W_{k}^{l}, b_{k}^{l}\} \forall l,k$, following gradients are to be computed: $\frac{\partial E}{\partial w_{k}^{l}}$, $\frac{\partial E}{\partial b_{k}^{l}}$, where $w_{k}^{l}$ denotes an arbitrary scalar entry in filter $W_{k}^{l}$. For simplicity, let output image $Y$ be of dimension $N\times N$. The equation for computing the gradient of weight $w_{k}^{l}$ in layer $l$ is given by:
	\begin{equation}\label{eqW3}
	\frac{\partial E}{\partial w_{k}^{l}} = -(Y_{g} - Y)\diamond\frac{\partial Y}{\partial w_{k}^{l}} + \alpha D_{R_{\delta}}\diamond\frac{\partial Y}{\partial w_{k}^{l}} - \beta D_{V}\diamond\frac{\partial Y}{\partial w_{k}^{l}}\vspace{-2pt}
	\end{equation}
	%\begin{equation}\label{eqW2}
	%\frac{\partial l}{\partial W^{2}} = -(Y_{g} - Y).\frac{\partial Y}{\partial W^{2}} + D_{R_{\delta}}.\frac{\partial Y}{\partial W^{2}} + D_{V}.\frac{\partial Y}{\partial W^{2}}\vspace{-6pt}
	%\end{equation}
	%\begin{equation}\label{eqW1}
	%\frac{\partial l}{\partial W^{1}} = -(Y_{g} - Y).\frac{\partial Y}{\partial W^{1}} + D_{R_{\delta}}.\frac{\partial Y}{\partial W^{1}} + D_{V}.\frac{\partial Y}{\partial W^{1}}\vspace{-6pt}
	%\end{equation}
	where $\diamond$ between two matrices $A$ and $B$ is defined as $\sum_{i,j}A_{i,j}B_{i,j}$, $D_{R_{\delta}} = -U\mbox{diag}\Bigg(-\frac{\sigma_{1}}{\delta^{2}}e^{-\sigma_{1}^{2}/2\delta^{2}},\ldots,-\frac{\sigma_{R}}{\delta^{2}}e^{-\sigma_{R}^{2}/2\delta^{2}}\Bigg)Z^{T}$ is the gradient of $R_{\delta}(Y)$ and $D_{V}$ is the gradient for $V(Y)$. The complete expression for $D_{V}$ is given by:
	\begin{equation*}\resizebox{\linewidth}{!}{$
		D_{V} = [v_{i,j}], \mbox{  }v_{i,j} = d_{i,j} - \frac{1}{4}(d_{i-1,j} + d_{i+1,j} + d_{i, j-1} + d_{i, j+1})$},
	\end{equation*}
	\begin{equation*}\resizebox{\linewidth}{!}{$
		d_{i,j} = \frac{2}{(N^{2})(N^{2} - 1)}\big(N^{2}p_{i,j} - \sum_{a}\sum_{b}p_{a,b} - \sum_{m}\sum_{n}(p_{m,n} - \frac{\sum_{a}\sum_{b}p_{a,b}}{N^{2}})\big)$},
	\end{equation*}
	where $P = [p_{i,j}]$, and $P$ is obtained by convolving $Y$ with a $3\times 3$ laplacian operator $L$. Expression for $p_{i,j}$ is given by:
	\begin{equation*}\resizebox{\linewidth}{!}{$\label{eqDv3}
		p_{i,j} = y_{i,j} - \frac{1}{4}(y_{i-1,j} + y_{i+1,j} + y_{i, j-1} + y_{i, j+1}), \mbox{ and } Y = [y_{i,j}]
		$}\end{equation*}
	Detailed derivations for the above equations are reported in the Appendix. Note that the gradient for bias terms are also updated in a similar fashion. The partial derivative $\frac{\partial Y}{\partial {w_{k}}^{l}}$ is obtained by a standard back propagation rule \cite{lecun1998gradient, werbos1994roots}.
	
	\section{DNSP with Data Adaptive Sharpness Enhancing Filters}
	\label{sec:LALF}
\begin{figure*}
		\centering
		\begin{center}
			\includegraphics[width=\linewidth]{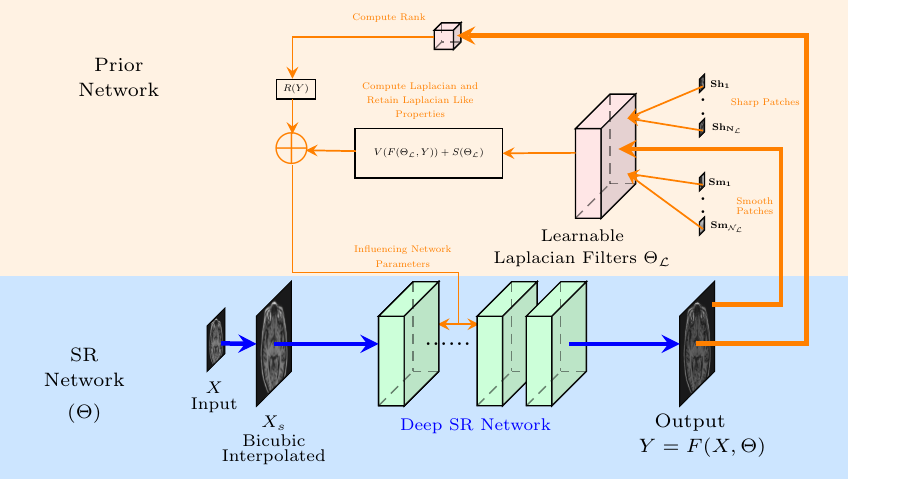}
		\end{center}
		%\vspace{-.5cm}
		\caption{The proposed DNSP architecture with learnable, data-adaptive sharpness. Bottom part of the network may be a typical state of the art LR to HR mapping, i.e.\ the SR network. As in Fig. 4, the parameters of the SR network are guided by prior information, except that in this case the sharpness filters are jointly learned with the network parameters $\Theta$. Post learning, i.e. during inference, the SR network carries out the LR to HR mapping.}
		
		\label{fig:DNSP_MOD}
	\end{figure*}

	%However, a fixed laplacian filter is more sensitive to noise and might also enhance some spurious components in an image. To counter this, we propose to use a learnable sharpness enhancing layer which is discussed with more details in Section \ref{sec:LALF}. Before going into the details of the learnable layer, we show the influence of priors to the parameters of the deep network via backpropagation. 
	A fixed Lapalacian filter can be sensitive to noise, enhance spurious components or might not be the best choice for a particular given data. Further, in the literature there exists a variety of sharpness enhancing filters \cite{sonka2014image} for different applications. Recall in Fig. \ref{fig:Net} that the Laplacian is computed through a fixed $3\times 3$ convolutional filter. To develop data-adaptive filters, we intend to learn a set of sharpness enhancing filters jointly with the network parameters instead of using a fixed Laplacian filter. For this purpose, the fixed laplacian layer in Figure \ref{fig:Net} is replaced by a set of filters that are initialized with the standard laplacian filter $L = [[0 \mbox{ }- 1 \mbox{ }0]^{T} [-1 \mbox{   4 } \mbox{ }-1]^{T} [0\mbox{ } - 1\mbox{ } 0]^{T}]$ with additive minor perturbations generated by a normal random variable with 0 mean and a small variance of $\epsilon = .0001$. The output of the SR network is passed through these filters which is followed by computing the average variance obtained from outputs of all these learnable sharpness filters. The extended architecture of our network that implements a learnable sharpness layer is shown in Fig. \ref{fig:DNSP_MOD}.

\begin{figure}[t]
		\begin{center}
			\includegraphics[width=\linewidth]{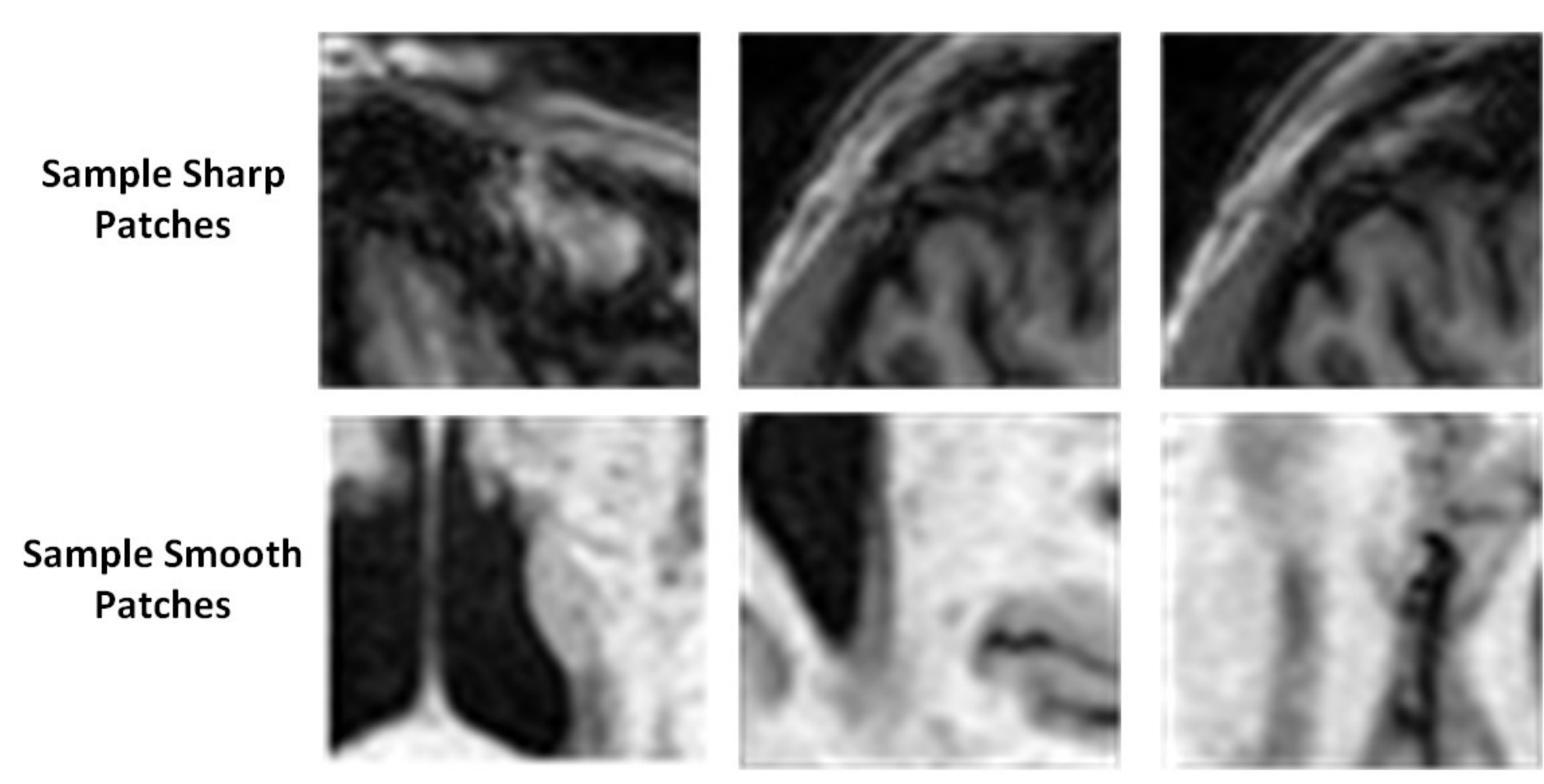}
		\end{center}
		%\vspace{-.5cm}
		\caption{\small{Representative examples of sharp and smooth patches. Top row represents sharp patches and bottom row represents smooth patches}}
		\label{fig:sharpSmooth}
	\end{figure}

	We represent these filters by $\Theta_{\mathcal{L}} = \{W_{\mathcal{L}}^{i}\}_{i=1}^{N_{\mathcal{L}}}$, where $N_{\mathcal{L}}$ is the total number of learnable sharpness filters. The modified equation to compute the variance (sharpness prior) is:
	\begin{equation}
	V(F(\Theta_{\mathcal{L}}, Y)) = V_{mod}(Y) = \frac{1}{N_{\mathcal{L}}}\sum_{i=1}^{N_{\mathcal{L}}}var(W_{\mathcal{L}}^{i}\circledast Y)
	\end{equation}
	A first step towards learning data-adaptive filters is the selection of sharp and smooth training patches which we carry out as follows:
	\begin{itemize}
		\item From each training image, we extract two patches of size $P\times P$ of which one is sharpest and the other is smoothest.
		\item To find the sharpest patch $\{Sh_{i}\}_{i=1}^{Q}$ from the $i^{th}$ training image where $i\in{1,\dots,Q}$, we pass all the patches of size $P\times P$ through a standard laplacian filter and select the patch that gives the maximum response in the sense of Frobenious norm $\parallel \bullet \parallel_{F}^{2}$ of the patches.
		\item Similarly, to find the smoothest patch $\{Sm_{i}\}_{i=1}^{Q}$, select the patch that gives the minimum response in the sense of Frobenious norm $\parallel \bullet \parallel_{F}^{2}$ of the patches.
		\item A visual inspection of all the candidate $Q$ smooth and sharp patches is performed to arrive at $N_{T} \leq Q$ selected smooth and sharp patches. 
	\end{itemize}
	Note that a sharpness enhancing filter $W_{\mathcal{L}}^{i}$ is expected to give the maximum response for the sharpest patches and the minimum response for the smoothest patches. This behavior is captured by formulating the following regularization term:
	\begin{equation}
	S(\Theta_{\mathcal{L}}) = \sum_{i=1}^{N_{\mathcal{L}}}\sum_{j=1}^{N_{T}}\parallel W_{\mathcal{L}}^{i}\circledast Sm_{j}\parallel_{F}^{2} - \sum_{i=1}^{N_{\mathcal{L}}}\sum_{j=1}^{N_{T}}\parallel W_{\mathcal{L}}^{i}\circledast Sh_{j}\parallel_{F}^{2}
	\end{equation}
	A negative sign before the response of sharp patches indicates that we intend to maximize it. Figure \ref{fig:sharpSmooth} shows three representative examples of sharp and smooth patches extracted via the procedure we discussed above. The new regularized loss function to learn the parameters of the SR network and data-adaptive sharpness filters is given by:
	\begin{align}
	\label{eq:loss_mod}
	E_{mod}(\Theta) &= \underbrace{\frac{1}{2}\|Y_{g} - F(X_{s}, \Theta)\|_{F}^{2}}_{MSE} + \underbrace{\alpha R_{\delta}(Y)}_{Low\mbox{ }Rank\mbox{ }Prior} \nonumber \\&- \underbrace{\beta V_{mod}(Y)}_{Sharpness\mbox{ }Prior} + \underbrace{\gamma S(\Theta_{\mathcal{L}})}_{Sharpness\mbox{ }Enhancing\mbox{ }Measure}
	\end{align}

	\textbf{Modified Back-Propagation Equations:} Extending Eq. (\ref{eqW3}), we obtain:  
	\begin{equation}\label{eq:bp_mod}
	\frac{\partial E_{mod}}{\partial w_{k}^{z}} = -(Y_{g} - Y)\diamond\frac{\partial Y}{\partial w_{k}^{z}} + \alpha D_{R_{\delta}}\diamond\frac{\partial Y}{\partial w_{k}^{z}} - \beta D_{V_{mod}}\diamond\frac{\partial Y}{\partial w_{k}^{z}}\vspace{-2pt}
	\end{equation}
	where $w_{k}^{z}$ is an arbitrary network parameter in layer $z$. Note that $w_{k}^{z}$ does not depend on $S(\Theta_{\mathcal{L}})$, which is not reflected in back-propagation equation. However, its influence is felt on learnable sharpness filters. The term $D_{R_{\delta}}$ remains the same as described in Section \ref{sec:DNSP_fix}. The complete expression for $D_{V_{mod}}$ is given by:
	\begin{equation}
	D_{V_{mod}} = [v_{i,j}], \mbox{  }v_{i,j} = \frac{1}{N_{\mathcal{L}}}\sum_{l=1}^{N_{\mathcal{L}}}\sum_{k,m=-1}^{1}W_{\mathcal{L}_{k,m}}^{l} d_{i-k,j-m}^{l} ,
	\end{equation}
	\begin{equation*}\resizebox{\linewidth}{!}{$
		d_{i,j}^{l} = \frac{2}{(N^{2})(N^{2} - 1)}\big(N^{2}p_{i,j}^{l} - \sum_{a}\sum_{b}p_{a,b}^{l} - \sum_{m}\sum_{n}(p_{m,n}^{l} - \frac{\sum_{a}\sum_{b}p_{a,b}^{l}}{N^{2}})\big)$},
	\end{equation*}
	where $P^{l} = [p_{i,j}^{l}]$, and $P^{l}$ is obtained by convolving $Y$ with the $l^{th}$ $3\times 3$ learnable sharpness filter $W_{\mathcal{L}}^{l}$. The expression for $p_{i,j}^{l}$ is given by:
	\begin{equation*}
	p_{i,j}^{l} = \sum_{k,m=-1}^{1}W_{\mathcal{L}_{k,m}}^{l} y_{i,j}, \mbox{ and } Y = [y_{i,j}]
	\end{equation*}
	The back-propagation equations for the sharpness filter parameter $W_{\mathcal{L}_{k,m}}^{l}$ are given by:
	\begin{equation}\label{eq:bp_llap}
	\frac{\partial E_{mod}}{\partial W_{\mathcal{L}_{k,m}}^{l}} =  - \beta \frac{\partial V_{mod}(Y)}{\partial W_{\mathcal{L}_{k,m}}^{l}} + \gamma \frac{\partial S(\Theta_{\mathcal{L}})}{\partial W_{\mathcal{L}_{k,m}}^{l}}
	\end{equation}
	where $W_{\mathcal{L}_{k,m}}^{l}$ is the $(k,m)$ coefficient in $l^{th}$ learnable filter $W_{\mathcal{L}}^{l}$ and $(k,m)\in \{-1,0,1\}$. Note that gradient of $W_{\mathcal{L}_{k,m}}^{l}$ is not dependent on the first two terms of the loss function in Eq. (\ref{eq:loss_mod}). $\frac{\partial S(\Theta_{\mathcal{L}})}{\partial W_{\mathcal{L}_{k,m}}^{l}}$ is given by:
	\begin{equation}
	\frac{\partial S(\Theta_{\mathcal{L}})}{\partial W_{\mathcal{L}_{k,m}}^{l}} = 2 \sum_{a=1}^{N_{T}}(W_{\mathcal{L}}^{l}\circledast Sm_{a})\diamond Sm_{a}' - 2 \sum_{a=1}^{N_{T}}(W_{\mathcal{L}}^{l}\circledast Sh_{a})\diamond Sh_{a}'
	\end{equation}
	where $Sm_{a}' = [s_{a_{i-k,j-m}}]$, $s_{a_{i-k,j-m}}$ is the $(i-k,j-m)$ coefficient of $Sm$. $Sh_{a}'$ can also defined in the similar fashion. $\frac{\partial V_{mod}(Y)}{\partial W_{\mathcal{L}_{k,m}}^{l}}$ is given by:
	\begin{equation}
	\frac{\partial V_{mod}(Y)}{\partial W_{\mathcal{L}_{k,m}}^{l}} = D_{V_{mod}}\diamond Y'
	\end{equation}
	where $Y' = [y_{i-k,j-m}]$ and $D_{V_{mod}}$ is the same matrix as defined previously. From equations (\ref{eq:bp_mod}) and (\ref{eq:bp_llap}), we can observe that weights of the learnable sharpness filter influence the SR network parameters and vice-versa. 
	
	\section{Experimental Evaluation}
	\label{sec:Experiments}
	%\vspace{-.3cm}
	\label{sec:Results}
	
	\subsection{Experimental Setup}
	\label{sec:exper_setup}
	
\begin{figure}[t]
		\begin{center}
			\includegraphics[width=.8\linewidth]{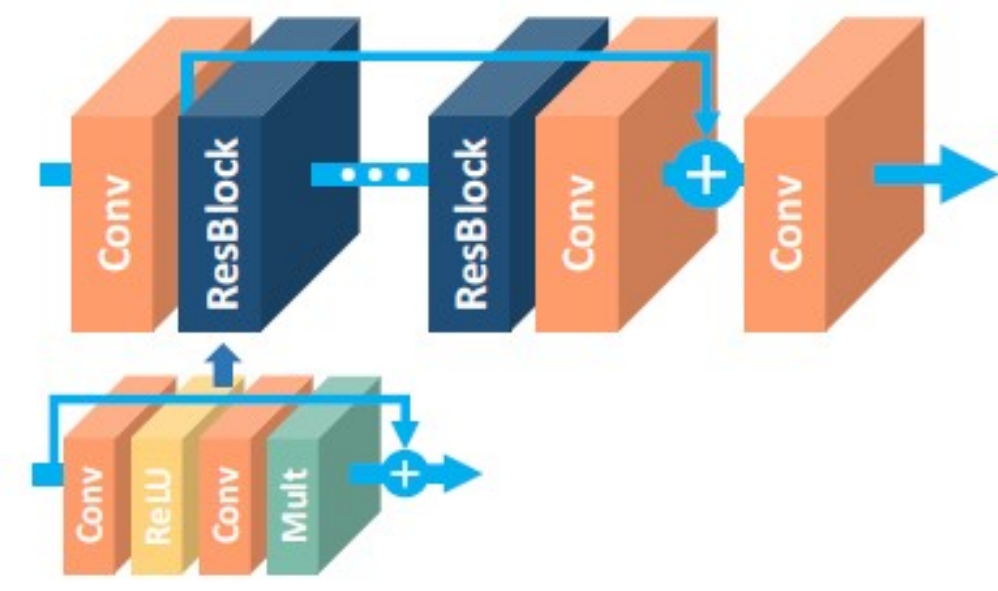}
		\end{center}
		\caption{\small{Illustration of EDSR architecture}}  
		\label{fig:EDSR_archi}
	\end{figure}

	\noindent \textbf{Databases, Training and Test Set-Up:} We evaluate the proposed DNSP on two publicly available MR brain image databases. The first database is 20 simulated T1 brain image stacks from Brainweb (BW)\footnote{\url{http://brainweb.bic.mni.mcgill.ca/brainweb/}}. Axial slices of these 20 stacks are distributed evenly for training and evaluation purposes. From each stack, we extract 40 slices making a total of 400 images for training and 400 images for evaluation. The second database we work with is from the Alzheimer's Disease Neuroimaging Initiative (ADNI)\footnote{\url{http://adni.loni.usc.edu/}}. The same training and test  configuration is employed as that of the BW database.\\
	\textbf{LR image simulation: } Consistent with \cite{trinh2014novel, shi2015lrtv}, we simulate training LR images by applying a gaussian blur and factor of $2$ downsampling. These LR images are then upscaled by bicubic interpolation. To speed up the training process, we further extract patches of size $40\times 40$ from these bicubic enlarged LR training images. Note that this is also a standard procedure used for training a typical deep SR network \cite{dong2016image, kim2016accurate,guo2017deep}.\\
	\textbf{Parameter Choices:} To obtain an accurate rank surrogate of $Y$, we  chose $\delta = .01$ based on guidelines mentioned in \cite{malek2014recovery}.
	We determine regularization weights in Eq. (\ref{eq:loss_mod}) as $\alpha=1\times 10^{-5}$, $\beta = 5\times 10^{-3}$, and $\gamma = 1\times 10^{-7}$ based on cross validation. More details can be found in the supplementary document. The number of learnable sharpness filters\footnote{It is observed that choosing $N_{\mathcal{L}} > 8$ did not offer any observable practical gains.} $N_{\mathcal{L}}$ is chosen as 8. The size of smooth and sharp patches $P$ is chosen as $40$ with the help of a domain expert. Batch size and number of epochs are chosen to be 64 and 50 for all the experiments.  \color{black} A total of around 6400 patches are extracted for ADNI dataset and 8000 patches are extracted for BW dataset. Therefore, 100 iterations are required to complete an epoch in the ADNI dataset and 125 iterations are required to complete an epoch in the BW dataset. \color{black} For optimization, an Adam optimizer \cite{kingma2014adam} with a learning rate of $1\times 10^{-4}$ is used. These values are consistent with other deep learning based SR methods \cite{dong2016image, chen2018brain, guo2017deep}.\\
	\textbf{Methods and Metrics for Comparison: }Two standard metrics PSNR and structural similarity index (SSIM) \cite{brunet2017optimizing} are used for evaluation. We compare  against following \textbf{six} methods: 
	
	\begin{itemize}
		\item Bicubic interpolation (BC)\cite{lehmann1999survey}- a fast baseline method. 
		\item  SRSW \cite{trinh2014novel} - an example based SR via sparse weighting (SRSW) for medical image SR, represents a state-of-the-art sparsity based method published in 2014. 
		\item LRTV \cite{shi2015lrtv}- amongst the most competitive model based approaches, involves low-rank (nuclear norm) and total variation (LRTV) regularizers published in 2015.
		\item SRCNN \cite{dong2016image}- the most widely used deep SR network, published in 2016.  
		\item EDSR \cite{lim2017enhanced} - a recent state of the art network for SR. Enhanced Deep Super-Resolution network (EDSR) won the first place in NTIRE 2017 competition \cite{timofte2017ntire}.
		\item DCSRN \cite{chen2018brain}-  Densely Connected Super-Resolution Network (DCSRN), a state-of-the-art deep learning approach developed specifically for MR image super-resolution, published in 2018. 
	\end{itemize}
	\textbf{Network Architecture:} As mentioned previously, the two proposed priors can be integrated into any deep SR network. Two deep SR networks we used are SRCNN and EDSR. SRCNN architecture is illustrated in Fig. \ref{fig:Net}. Figure \ref{fig:EDSR_archi} illustrates the architecture of EDSR. It is composed of total $32$ residual blocks wherein a convolutional layer of a residual block consists of $256$ $3\times 3\times 256$ filters. First layer is composed of 256 $3\times 3 \times 1$ filters and last layer consists of one $3\times 3\times 256$ filter. A detailed description of network architectures for both the methods can be found in \cite{dong2016image, lim2017enhanced}.  \color{black} Note that in the original EDSR architecture, the interpolation block is placed at the end of the residual network. In this work, to be consistent with SRCNN, we perform a bicubic interpolation prior to sending the input to EDSR network thereby removing the interpolation block after the residual network. Hence the size of the input send to the EDSR network is same as the size of the desired output. We did not observe noticeable performance difference by shifting the interpolation block, hence we chose the configuration that is consistent with other deep learning frameworks. \color{black}More details can be found in the supplementary document. \color{black} Unless otherwise stated, note that our priors are integrated with the EDSR network.
	\textbf{Remark:} Note that our choice of SRCNN and EDSR as base DNSP networks is because SRCNN is widely used and EDSR has recently been shown to be one of the best performing methods (winner of the 2017 NTIRE contest at IEEE conference on Computer Vision and Pattern recognition (CVPR)). The goal in this work is to demonstrate the value of priors in enhancing performance and not to perform an exhaustive comparison of deep SR architectures \cite{timofte2017ntire,timofte2018ntire}.   
	
	\subsection{Significance of Priors: DNSP Variants}
	\label{sec:res}
	
\begin{figure}[t]
		\begin{center}
			\includegraphics[width=\linewidth]{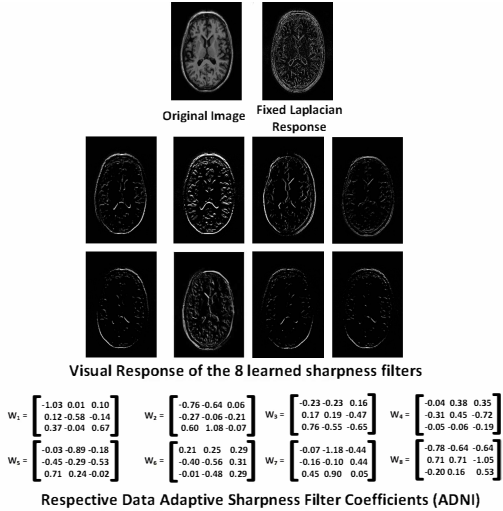}
		\end{center}\captionsetup{labelfont={color=black},font={color=black}}
		\caption{\small{Response of the 8 learned sharpness filters (along with the corresponding filter coefficients) vs. the Laplacian filter on a sample image from the ADNI dataset, as well as the coefficients of the learned filters.}}  
		\label{fig:vis_lap}
	\end{figure}

	In this section, we report the results for different variants of the proposed DNSP to bring out the value added by each prior. We name the variants as follows: 1) DNSP-NP, network with no priors which is same as EDSR, 2) DNSP-LR, network with only low rank prior, 3) DNSP-FS, network with only sharpness prior with a fixed Laplacian layer, 4) DNSP-LS, network with only sharpness prior but with learnable sharpness filters and finally 5) DNSP-AP, network with both the priors included along with learnable sharpness filters. Table \ref{tab:results_varian} shows the PSNR and SSIM on both the datasets. We can observe that priors improve the performance of the network. Among the individual priors, we observe the best performance for DNSP-LS, which is expected as the sharpness is enhanced via a data adaptive procedure that exploits available training. Figure \ref{fig:vis_lap} shows a comparison of the response from fixed laplacian filter and $N_{\mathcal{L}} = 8$ filters that are learned via DNSP-LS method. We can observe that spurious (undesirable/noise-like) edges that are present in the fixed Laplacian response are minimally seen in the response of the 8 filters learned based on data, which on the other hand lead to sharper images overall. It can be observed that the responses of learned sharpness filter depart from that of the Laplacian 
	( for example in exhibiting some directional orientation), which is a result of training image data adaptation.   
	
\begin{figure*}[t!]
		\centering
		%	\minipage{0.5\textwidth}
		%	\centering
		%	\includegraphics[width=\linewidth]{validations_ADNI.png}
		%	(a) ADNI
		%	\endminipage\hfill
		%	\minipage{0.5\textwidth}
		%	\centering
		%	\includegraphics[width=\linewidth]{validations_BW.png}
		%	(b) BW
		%	\endminipage
		\includegraphics[width=0.8\linewidth]{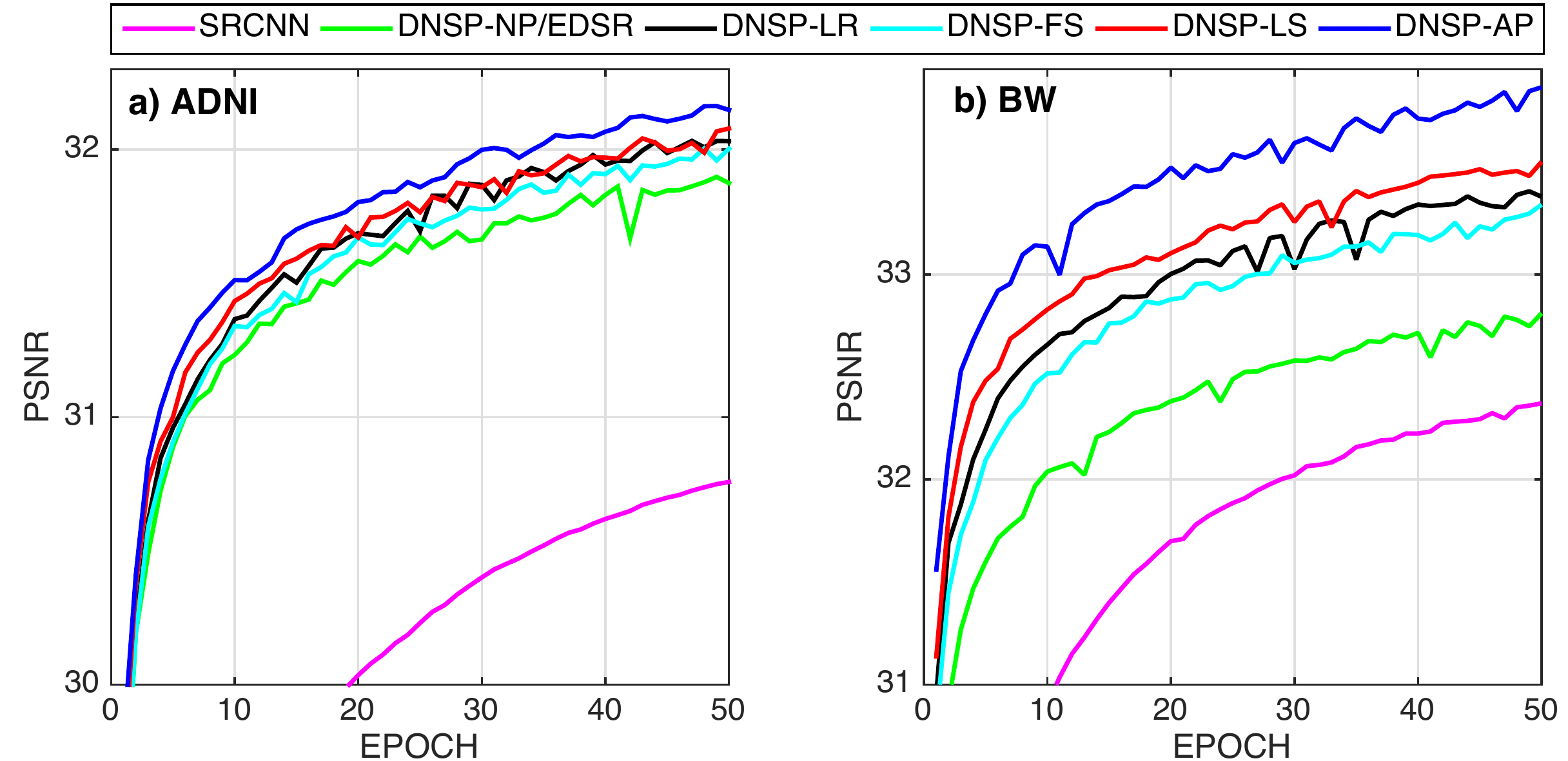}\captionsetup{labelfont={color=black},font={color=black}}
		\caption{Validation curves for different variations of our DNSP method on a)ADNI, and b)BW datasets.  SRCNN and EDSR results are also included. }
		\label{fig:validations}
	\end{figure*}

	 \color{black}Further, to provide more insights about the performance of different priors, validation curves for PSNR vs EPOCH on test sets are illustrated in Fig. \ref{fig:validations}. It can be observed that the network with priors always outperforms the one without priors. As expected DNSP-AP is the best performing method. We also observe that DNSP-LS does better than DNSP-FS and DNSP-LR. \color{black}For all the subsequent experiments, unless otherwise stated, we report results of DNSP-AP.  
	%Note that, we did not report results for the scenario of all priors but with fixed laplacian layer. This is due to the fact that DNSP-LS does better than DNSP-FS.
	
	\begin{table}[h]
		\caption{\small{PSNR and SSIM comparisons for different regularizers}}
		\label{tab:results_varian}
		%\vspace{-.5cm}
		\begin{center}
			\resizebox{.6\linewidth}{!}
			{\begin{tabular}{cccc}
					\hline\hline
					\textbf{Method} & \textbf{Database} & \textbf{PSNR} & \textbf{SSIM}\\
					\hline
					\multirow{2}{*}{DNSP-NP} & BW & $32.8112$ & $.8803$ \\
					& ADNI  & $31.8945$ & $.9503$ \\
					\hline
					\multirow{2}{*}{DNSP-LR} & BW & $33.3801$ & $.8835$ \\
					& ADNI  & $32.03$ & $.9504$ \\
					\hline
					\multirow{2}{*}{DNSP-FS} & BW & $33.3466$ & $.8818$ \\
					& ADNI  & $32$ & $.9525$ \\
					\hline
					\multirow{2}{*}{DNSP-LS} & BW &$33.54$ & $.8871$ \\
					& ADNI  & $32.0825$ & $.9530$ \\
					\hline
					\multirow{2}{*}{DNSP-AP} & BW &$\textbf{33.9170}$ & $\textbf{.8902}$ \\
					& ADNI  & $\textbf{32.1364}$ & $\textbf{.9534}$ \\
					\hline
			\end{tabular}}
		\end{center}
	\end{table}

	\begin{table*}[h]
		\caption{\small{PSNR and SSIM comparisons against competing methods for different scaling factors}}
		\label{tab:results}
		%\vspace{-.5cm}
		\begin{center}
			\resizebox{.8\linewidth}{!}
			{\begin{tabular}{cc|cc|cc|cc}
					\hline\hline
					\textbf{Method} & \textbf{Database} & \textbf{PSNR (x2)} & \textbf{SSIM (x2)}& \textbf{PSNR (x3)} & \textbf{SSIM (x3)}& \textbf{PSNR (x4)} & \textbf{SSIM (x4)}\\
					\hline
					\multirow{2}{*}{BC} & BW & $29.09$ & $.8369$ & $24.63$ & $.7516$ & $23.19$ & $.7211$\\
					& ADNI  & $27.82$ & $.8958$ & $23.19$ & $.8014$ & $21.96$ & $.7605$\\
					\hline
					\multirow{2}{*}{SRSW[\cite{trinh2014novel},TIP 2014]} & BW & $31.16$ & $.80$ &$26.66$ & $.7421$ &$23.83$ & $.7265$ \\
					& ADNI  & $30.19$ & $.7721$ & $25.04$ & $.7157$ & $22.78$ & $.6953$ \\
					\hline
					\multirow{2}{*}{LRTV[\cite{shi2015lrtv}, TMI 2015]} & BW & $30.46$ & $.8560$ & $26.21$ & $.7562$ & $23.76$ & $.7221$ \\
					& ADNI  & $30.50$ & $.7830$ & $25.22$ & $.7210$ & $22.31$ & $.7032$ \\
					\hline
					\multirow{2}{*}{SRCNN[\cite{dong2016image}, TPAMI 2016]} & BW &$32.37$ & $.8762$ &$28.48$ & $.7923$ &$25.00$ & $.7418$ \\
					& ADNI  & $30.75$ & $.9380$ & $26.85$ & $.8589$ & $24.26$ & $.7553$ \\
					\hline
					\multirow{2}{*}{DCSRN[\cite{chen2018brain}, ISBI 2018]} & BW &$32.63$ & $.8785$ &$29.03$ & $.8004$ &$25.31$ & $.7474$ \\
					& ADNI  & $31.06$ & $.9417$ & $27.16$ & $.8630$ & $24.50$ & $.7594$ \\
					\hline
					\multirow{2}{*}{EDSR[\cite{lim2017enhanced}, CVPR 2017]} & BW &$32.81$ & $.8803$ &$29.61$ & $.8156$ &$25.58$ & $.7668$ \\
					& ADNI  & $31.89$ & $.9503$ & $27.51$ & $.8753$ & $24.77$ & $.8280$ \\
					\hline
					\multirow{2}{*}{DNSP-SRCNN-AP} & BW &$32.76$ & $.8788$ &$29.26$ & $.8021$ &$25.43$ & $.7555$ \\
					& ADNI  & $31.27$ & $.9458$ & $27.40$ & $.8711$ & $24.52$ & $.8139$ \\
					\hline
					\multirow{2}{*}{DNSP-EDSR-AP} & BW &$\textbf{33.92}$ & $\textbf{.8902}$ &$\textbf{30.50}$ & $\textbf{.8312}$ & $\textbf{26.72}$ & $\textbf{.7742}$ \\
					& ADNI  & $\textbf{32.14}$ & $\textbf{.9534}$ & $\textbf{28.02}$ & $\textbf{.8757}$ & $\textbf{25.78}$ & $\textbf{.8253}$ \\
					\hline
			\end{tabular}}
		\end{center}
	\end{table*}
	
	\begin{figure}[t]
		\begin{center}
			\includegraphics[width=\linewidth]{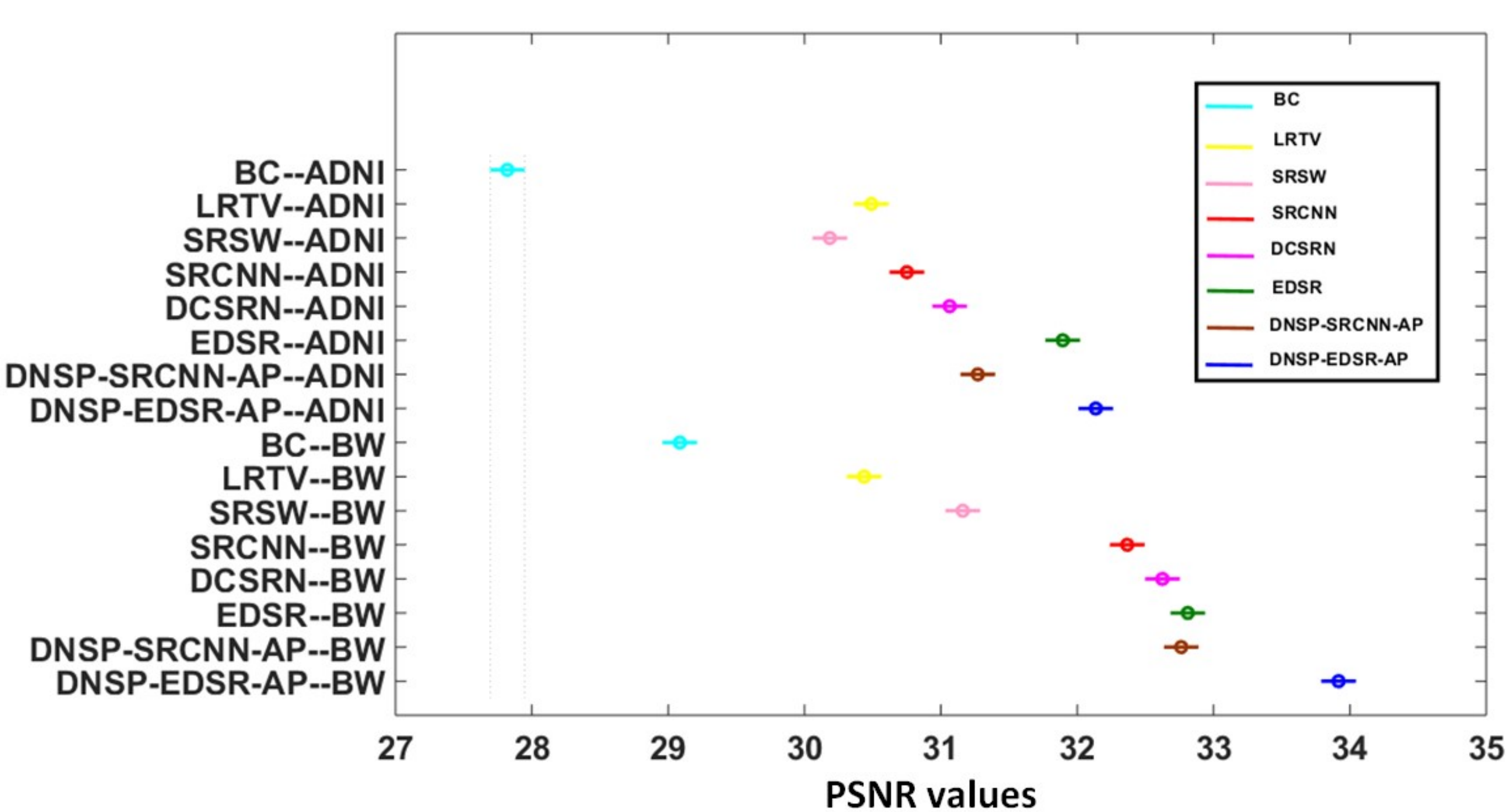}
		\end{center}
		\caption{\small{2-way ANOVA comparing DNSP vs. competing methods. The intervals represent the 95 $\%$ confidence intervals of PSNR values for a given configuration of method--dataset. Values reported for ANOVA across the method factor are $df = 7$, $F = 1466.94$, $p\ll .01$.}}
		\label{fig:ANOVA}
	\end{figure}
	
	\begin{figure*}[t]
		\begin{center}
			\includegraphics[width=\linewidth]{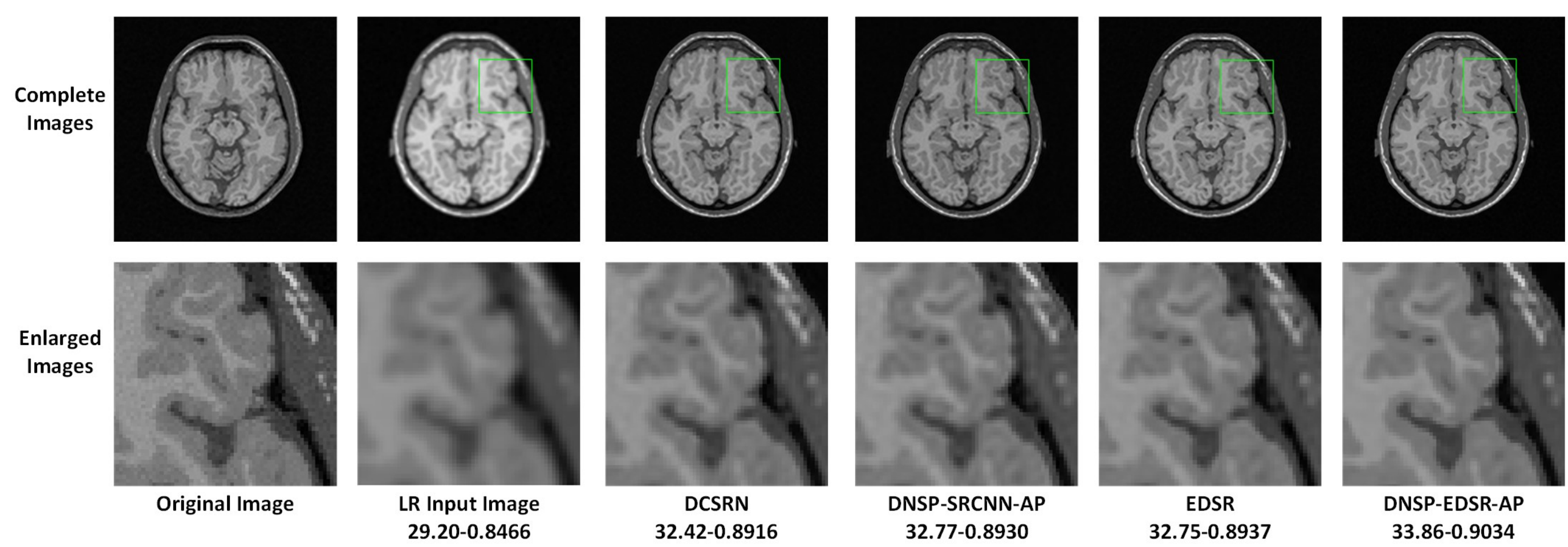}
		\end{center}
		\caption{\small{Comparisons of top 4 methods for an image in BW data set for scale factor of 2. A small portion of the images (marked by green box) in the first row is zoomed in and shown in second row. The numerical figures constitute the respective PSNR-SSIM values.}}  
		\label{fig:images}
	\end{figure*}
	
	\begin{figure*}[t]
		\begin{center}
			\includegraphics[width=\linewidth]{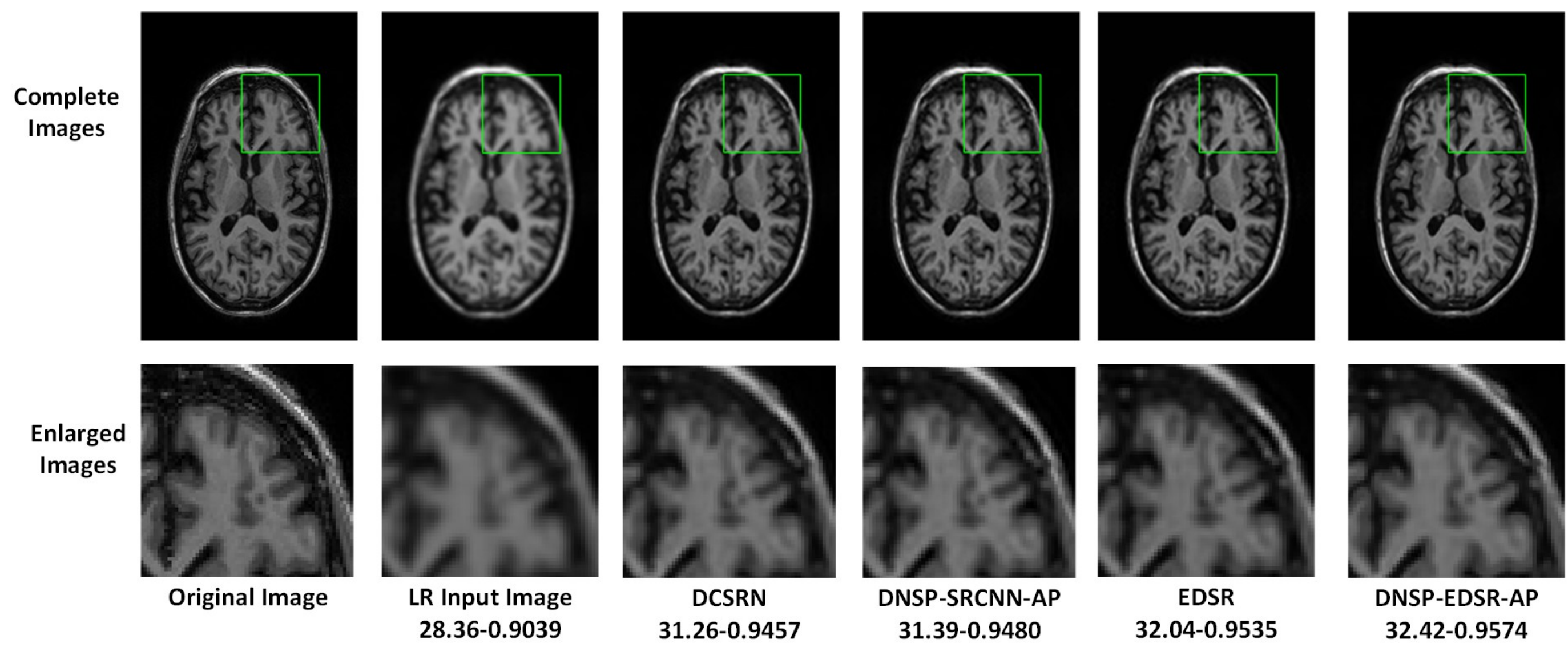}
		\end{center}
		\caption{\small{Comparisons of top 4 methods for an image in ADNI dataset for scale factor of 2. A small portion of the images (marked by green box) in the first row is zoomed in and shown in second row. The numerical figures constitute the respective PSNR-SSIM values.}}  
		\label{fig:images_ADNI}
	\end{figure*}

	\begin{figure*}[t]
	\begin{center}
		\includegraphics[width=.8\linewidth]{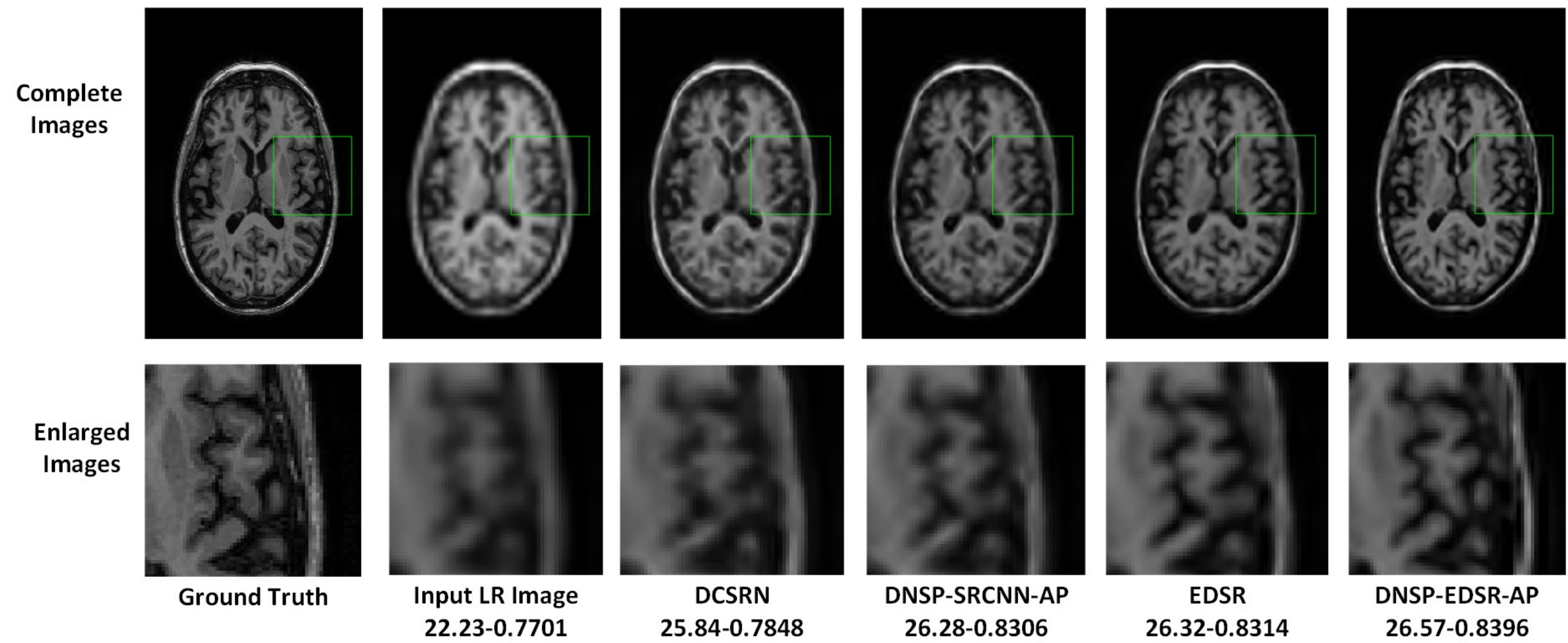}
	\end{center}\captionsetup{labelfont={color=black},font={color=black}}
	\caption{\small{Comparisons of top 4 methods for an image in ADNI dataset for scale factor of 4. A small portion of the images (marked by green box) in the first row is zoomed in and shown in second row. The numerical figures constitute the respective PSNR-SSIM values.}}  
	\label{fig:images_ADNI_4}
\end{figure*}

\begin{figure*}[t]
	\begin{center}
		\includegraphics[width=.8\linewidth]{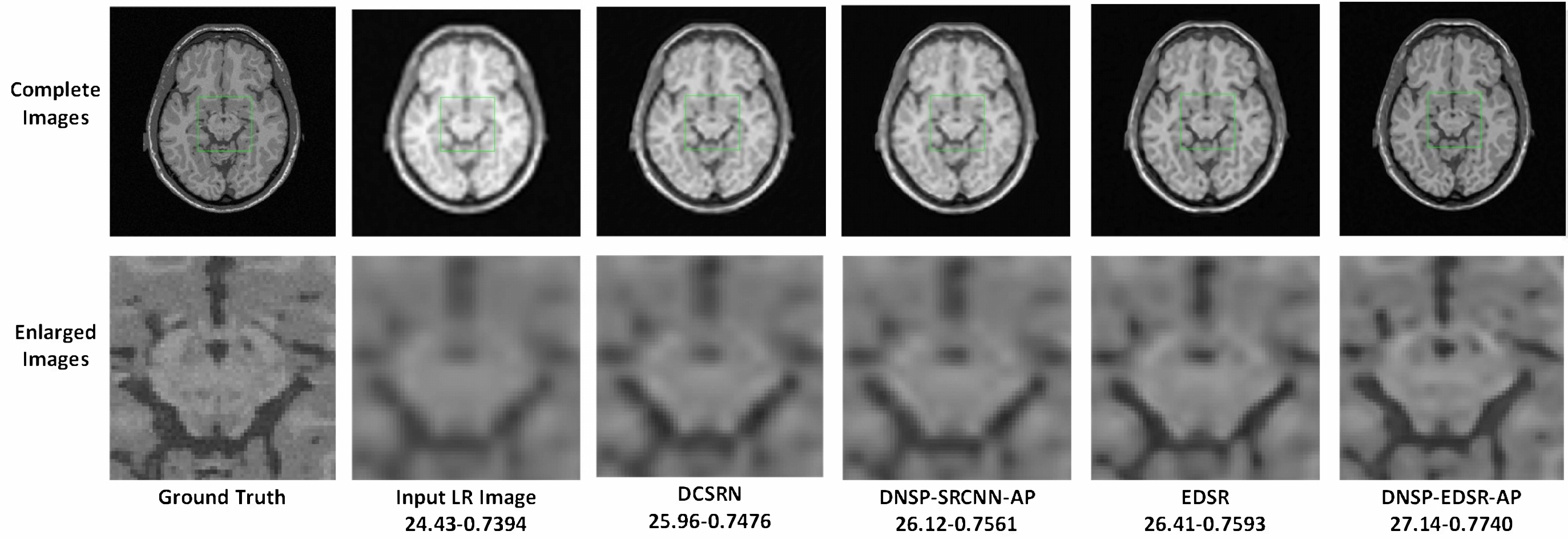}
	\end{center}\captionsetup{labelfont={color=black},font={color=black}}
	\caption{\small{Comparisons of top 4 methods for an image in BW data set for scale factor of 4. A small portion of the images (marked by green rectangle) in the first row is zoomed in and shown in second row. The numerical figures constitute the respective PSNR-SSIM values.}}  
	\label{fig:images_BW_4}
\end{figure*}

	\begin{figure*}[!t]
		\begin{center}
			\includegraphics[width=\linewidth]{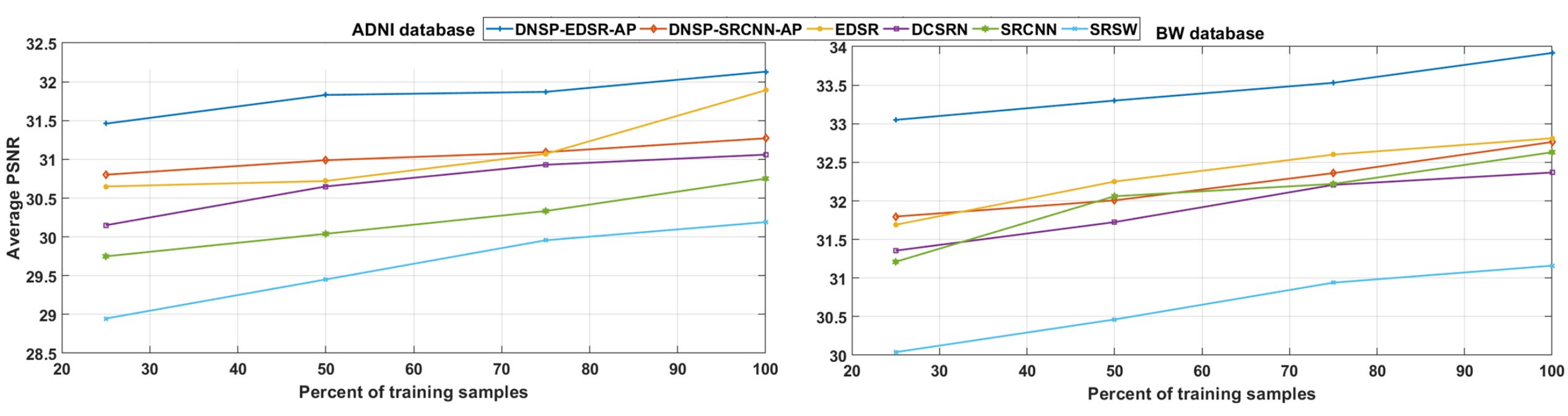}
		\end{center}
		\caption{\small{PSNR vs percent training samples.}}
		\label{fig:trainPlot}
	\end{figure*}
	
	\subsection{Comparisons Against State-of-the-Art Methods}
	\label{sec:comp} 
	
	Table \ref{tab:results} shows PSNR and SSIM values for all competing methods. Note that we used two different base networks for DNSP: 1) DNSP-SRCNN-AP - the base network is SRCNN and 2) DNSP-EDSR-AP - the base network is EDSR. Three trends emerge from the results: 1) DNSP-EDSR-AP outperforms the competition, 2) DNSP-SRCNN-AP does better than all the methods except EDSR, and 3) overall, deep SR methods, i.e.\ SRCNN, EDSR, DCSRN and DNSP perform better than other alternatives. To confirm this statistically, we performed a 2-way Analysis of Variance (ANOVA) on PSNR values for all the methods across the two datasets which is illustrated in Fig. \ref{fig:ANOVA}. It may be inferred from Fig. \ref{fig:ANOVA} that deep learning methods are statistically well separated from the traditional methods and further DNSP-EDSR-AP is well separated from all the competing methods indicating the effectiveness of using prior information. Figures \ref{fig:images} and \ref{fig:images_ADNI} illustrate the results of the top 4 methods w.r.t. PSNR on a sample image from BW and ADNI databases respectively  \color{black} for a down-sampling factor of 2  while Figures \ref{fig:images_ADNI_4} and \ref{fig:images_BW_4} show results for a down-sampling factor of 4. \color{black} DNSP-EDSR-AP particularly excels in recovering  fine image detail (enlarged with zoom-in boxes), thanks to data-adaptive sharpness. \\
	
	%We also observe that for limited training scenario, performance of SRCNN and SRSW is as low as LRTV which is not a learning based approach. Therefore, incorporating priors in a deep network enhances the performance effectively in a low training scenario which is the case for most of the medical imaging applications.

	%\vspace{-.5cm}
	\begin{figure}[t]
		\begin{center}
			\includegraphics[width=\linewidth]{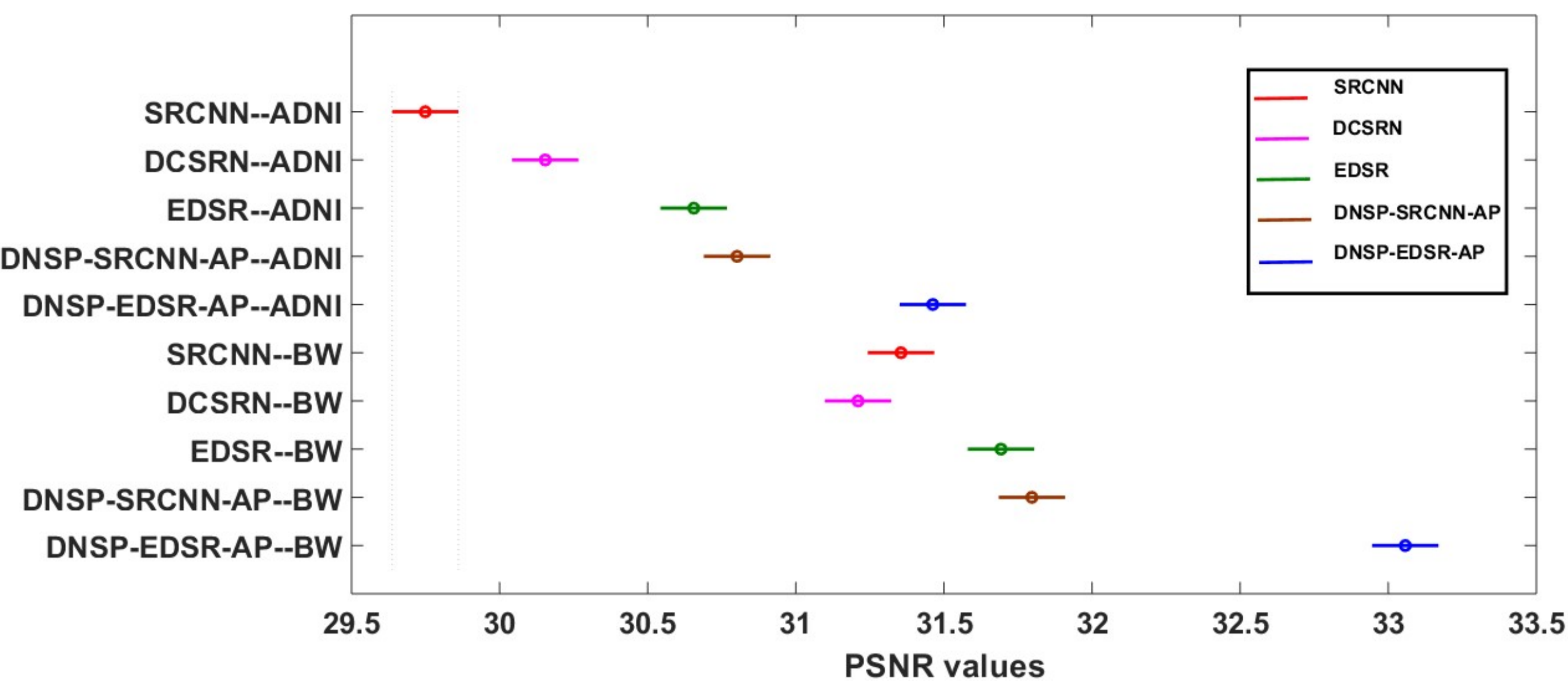}
		\end{center}
		%	\vspace{-.5cm}
		\caption{\small{2-way ANOVA comparing the deep learning methods for 25 percent training scenario. The intervals represent the 95 $\%$ confidence intervals of PSNR values for a given configuration of method--dataset. Values reported for ANOVA across the method factor are $df = 4$, $F = 362.23$, $p\ll .01$.}}
		\label{fig:ANOVA_low}
	\end{figure}

	%\vspace{-.5cm}
	
	\subsection{Performance in Varying Training Regimes}
	\label{sec:lowtrain}
	Figure \ref{fig:trainPlot} compares the performance of the learning based methods for different percentage of training samples considered on both the datasets. Twenty five, $50$ and $75$ percent of the 400 training images are employed. Two inferences can be made: 1) DNSP-EDSR-AP consistently outperforms  EDSR, SRCNN, DCSRN and SRSW, 2) The performance degradation of both DNSP-SRCNN-AP and DNSP-EDSR-AP is more graceful with a decrease in the number of training samples. For example, PSNR values for EDSR, SRCNN, and SRSW dropped by almost close to 1-1.5db whereas for DNSP-*, the drop is in between .5-1db, when the training drops to 25 percent. Another interesting observation is DNSP-SRCNN-AP does better than EDSR for the 25 percent training scenario. These results unequivocally demonstrate the value of priors (capturing domain specific signal structure) in enhancing performance when training imagery is limited.  \color{black}This is due to the fact that priors aid in capturing the structure of the images which in turn ensures that the deep network outputs images that are consistent with the structure of the original images. In the case of large training samples, the network has a sufficient amount of training samples to discover the inherent structure of the images. However, when the training data is limited, the network by itself does not have a sufficient amount of images to discover the inherent structure. In such cases priors guide the network to discover the appropriate structures of the underlying images, thereby enhancing the performance of the network. This fact is clearly brought out in Figure \ref{fig:validations} where validation curves of different variants of our proposed method are illustrated. There we can observe that the prior guided networks start performing better than DNSP-NP (EDSR with no priors) right from epoch 1 which confirms that the network is being guided to discover appropriate structures.  \color{black}

	Further, to confirm this statistically, Fig. \ref{fig:ANOVA_low} shows the 2-way ANOVA analysis for the 25 percent training scenario for all the deep learning based methods. It can be observed that DNSP-EDSR-AP is well separated from the other methods.  Note that LRTV and BC are excluded from this experiment since these methods are not learning based.

	\begin{figure*}[t]
		\begin{center}
			\includegraphics[width=.7\linewidth]{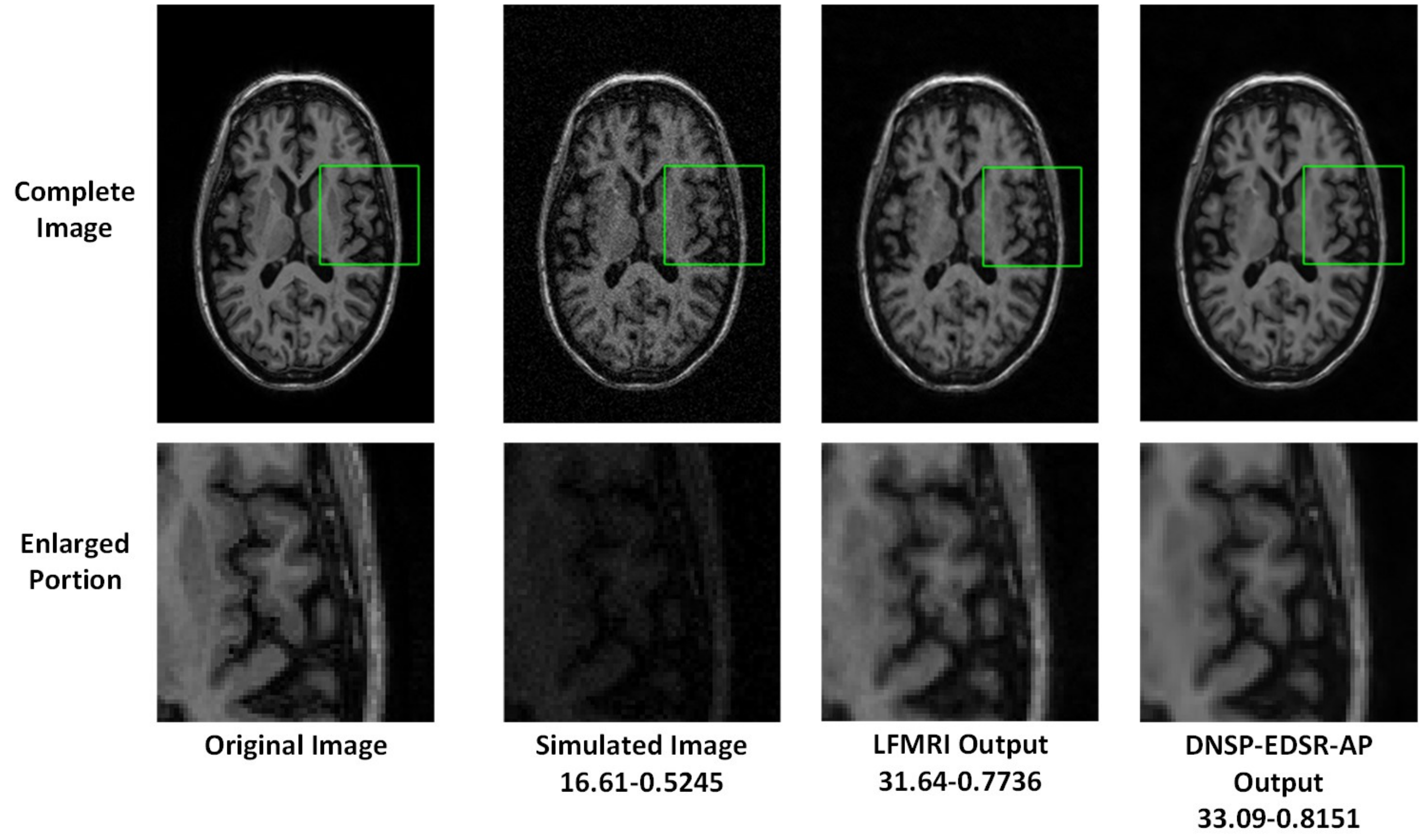}
			
		\end{center}
		\caption{\small{Comparisons of LFMRI and DNSP on low field MR images. The values shown constitute the respective PSNR-SSIM values.}} 
		\label{fig:images_lowfield}
	\end{figure*}
	
	\begin{figure}[!t]
		\begin{center}
			\includegraphics[width=\linewidth]{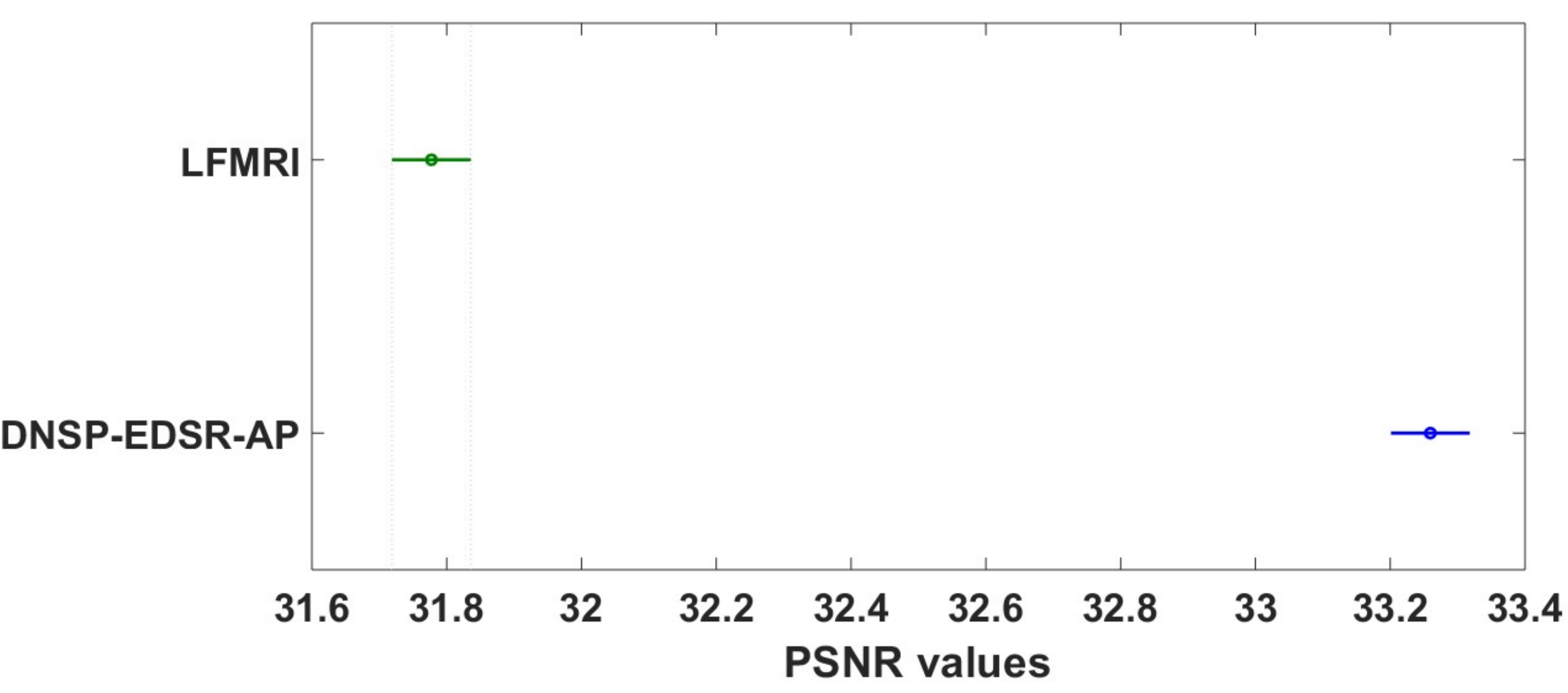}
		\end{center}
		%	\vspace{-.5cm}
		\caption{\small{1-way ANOVA comparing DNSP-AP vs. EDSR for low field simulated images. The intervals represent the 95 $\%$ confidence intervals of PSNR values for a given configuration of method-dataset. Values reported for ANOVA are $df = 1$, $F = 439.35$, $p\ll .01$.}}
		\label{fig:ANOVA_lowField}
	\end{figure}

	\begin{table}[t]
		\caption{\small{PSNR and SSIM comparisons for simulated low field MR images}}
		\label{tab:lowfield}
		\begin{center}
			\resizebox{.6\linewidth}{!}
			{\begin{tabular}{ccc}
					\hline\hline
					\textbf{Method} &  \textbf{PSNR} & \textbf{SSIM}\\
					\hline
					LFMRI &  $31.78$ & $.7760$ \\
					
					{DNSP-EDSR-AP} & $33.25$ & $.8205$ \\
					\hline

			\end{tabular}}
		\end{center}
	\end{table}
	
	\subsection{Enhancing low field MR images }
	\label{sec:low_field}
	
	A key practical task is discovering the mapping from low to high field MRI.  This is required in scenarios where expensive, high field MR captures are not available but low field MR images may be enhanced prior to diagnosis. This problem has received recent attention via learning based methods \cite{bahrami2016reconstruction,bahrami2016convolutional}. The dataset in \cite{bahrami2016reconstruction,bahrami2016convolutional} is however not publicly available. %and in general, to the best of our knowledge, publicly available databases that provide training images simultaneously from both high field and low field MR devices are hard to come by. 
	To circumvent this, we simulate low field MR images for the ADNI dataset using a recently developed technique in \cite{wu2016minimum}. Certain assumptions are made by the authors of \cite{wu2016minimum}: 1.) the noise model is assumed thermal, and 2.) Further, a single global relaxation correction function is used to account for the signal change at different field strengths. We point to \cite{wu2016minimum} for more details and the code that implements the degradation from high to low field MR. We use their code to simulate 1.5T images from the available 3T ADNI images. %Figure \ref{fig:sim} shows a representative example of 1.5T simulated image. 
	
	We report results on the 25 percent training setup as described before. For this experiment, we compared against \cite{bahrami2016convolutional} since it is also a deep learning method developed specifically for low to high field MR enhancement. We call this method as LFMRI. Table \ref{tab:lowfield} shows the results and the benefits of DNSP are readily apparent. Visual comparisons of enhanced images via both the methods are shown in Fig. \ref{fig:images_lowfield}. A one way statistical ANOVA is further performed (using 400 test images) to confirm that the benefits of DNSP are indeed statistically pronounced --  see Fig. \ref{fig:ANOVA_lowField}.  
	
	 \color{black}
	\subsection{Experiments on Real World Clinical image pairs}
	\label{sec:realistic}
\begin{figure*}[t]
		\begin{center}
			\includegraphics[width=.8\linewidth]{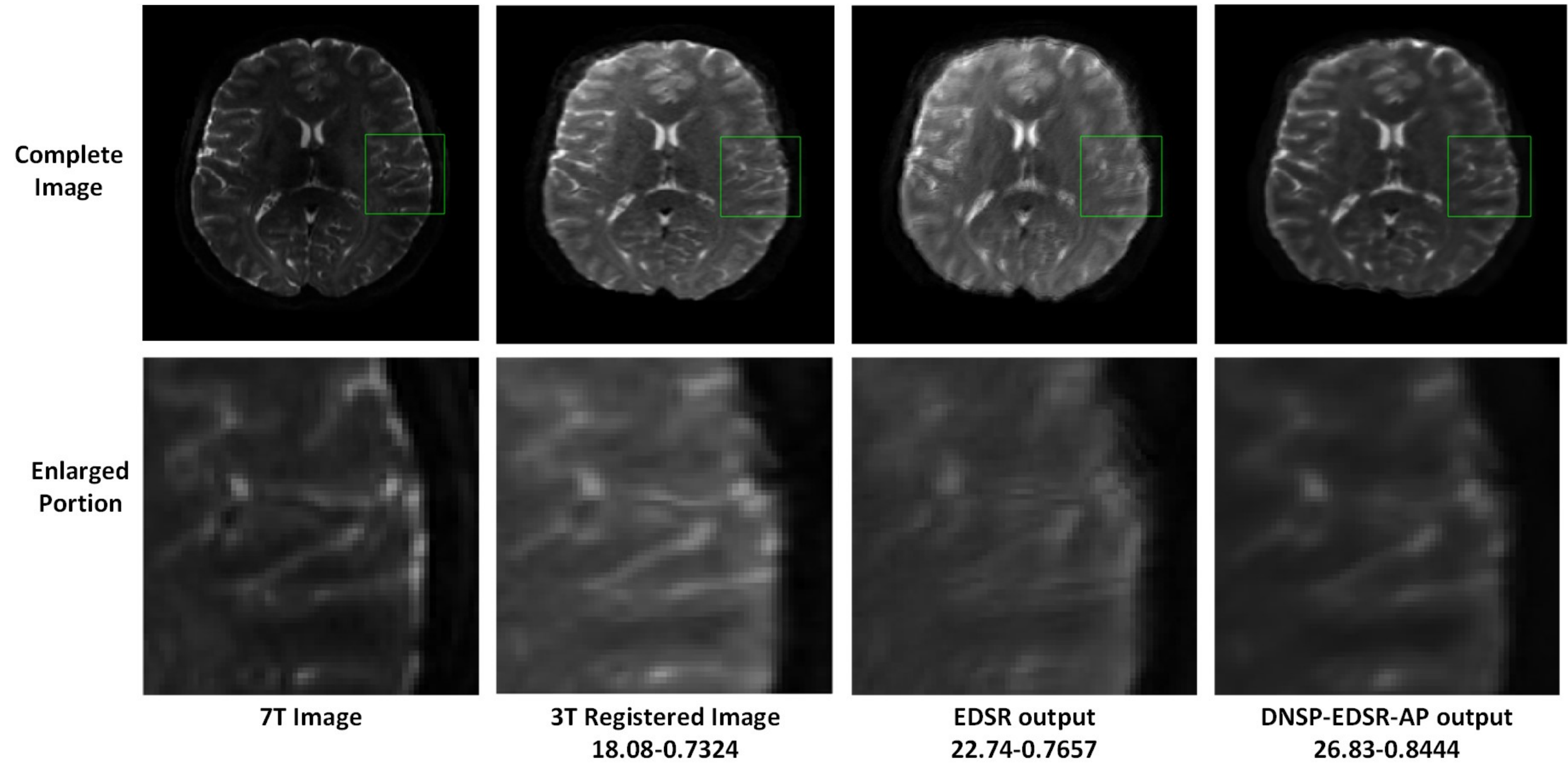}
			
		\end{center}
		\captionsetup{labelfont={color=black},font={color=black}}
		\caption{\small{Comparisons of EDSR and DNSP-EDSR-AP on 3T7T-DW dataset \cite{van2013wu}. The numerical assessment is shown as PSNR-SSIM. The DNSP-EDSR generates the best results both numerically and visually compared to EDSR for 3T7T-DW dataset.}} 
		\label{fig:3T7T}
	\end{figure*}

\begin{figure*}[t]
		\begin{center}
			\includegraphics[width=.8\linewidth]{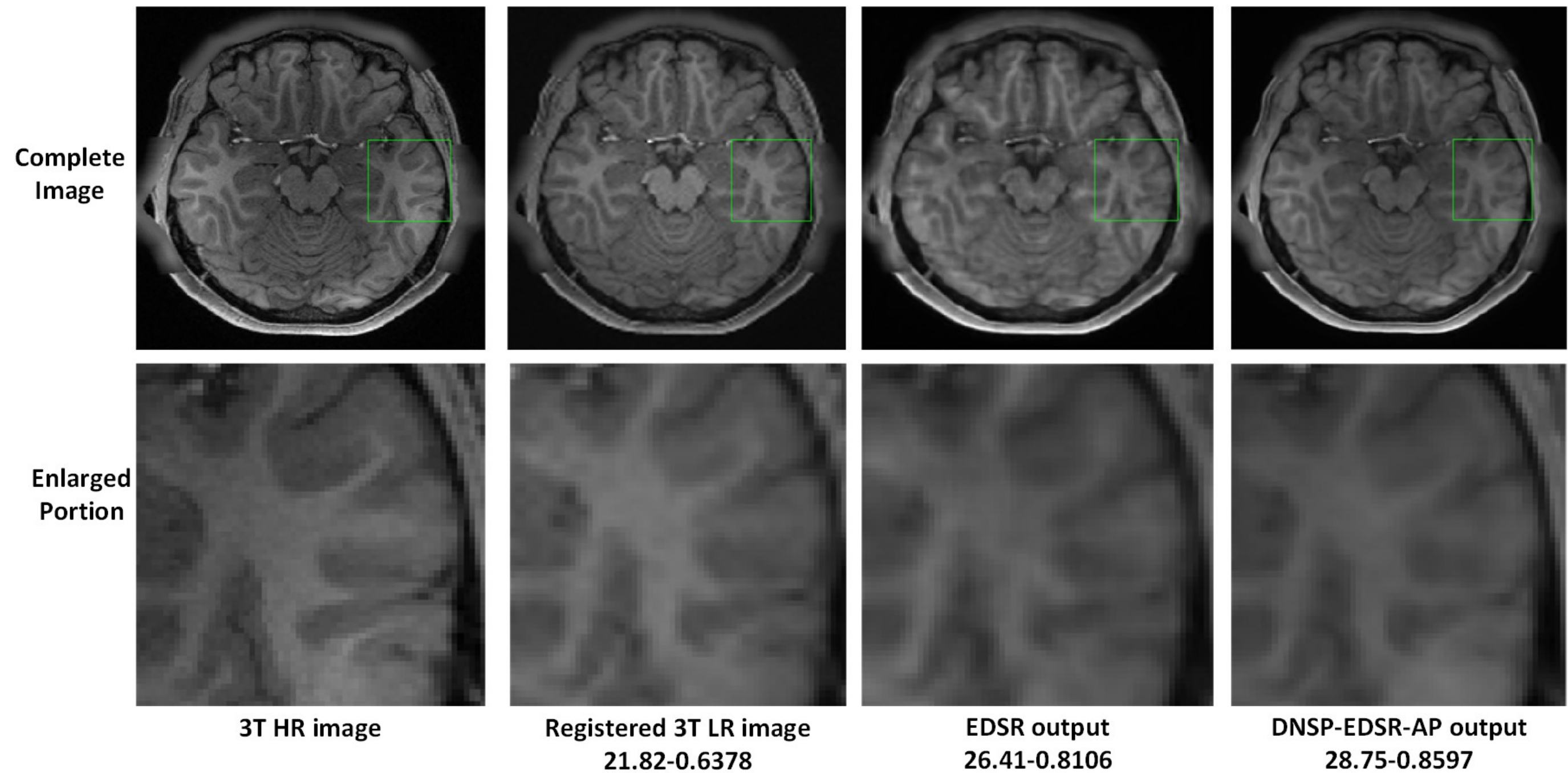}
			
		\end{center}
		\captionsetup{labelfont={color=black},font={color=black}}
		\caption{\small{Comparisons of EDSR and DNSP-EDSR-AP on 3T3T-T1 dataset \cite{van2013wu}. The numerical assessment is shown as PSNR-SSIM. The DNSP-EDSR generates the best results both numerically and visually compared to EDSR for 3T3T-T1 dataset.}} 
		\label{fig:3T3T}
	\end{figure*}

	To further validate our framework in real world scenarios, we perform experiments on two new datasets obtained from Human Connectome Project (HCP) \cite{van2013wu}\footnote{\url{https://db.humanconnectome.org/app/template/}}. Wide variety of datasets are available in the aforementioned link of which we selected two scenarios that are closely related to our work. 
	\begin{enumerate}
		\item 3T7T-DW: This scenario consists of the Diffusion Weighted (DW) MRI images of the same patients acquired at 3T (Tesla) and 7T magnetic field strengths. We extracted the data of 15 patients and used images from 5 patients for training and the images from the rest of 10 patients for testing. The same selection strategy is used 5 times and the results are averaged to remove the selection bias. All the 3T scans are obtained from a customized Siemens 3T “Connectome Skyra”. A Spin Echo sequence with a repetition time (TR) of 5520ms and an echo time (TE) of 89.5ms is used for acquisition. The dimension of the images in 3T are $168\times 144\times 111$ in axial plane. The 7T scans are acquired by a Siemens Magnetom 7T MR scanner. A Spin Echo sequence with a TR of 7000ms and a TE of 71.2ms is used for acquisition. The dimension of the images in 7T are $200\times 200\times 132$ in axial plane. Before extracting the patches for training as described in Section \ref{sec:exper_setup}, we perform registration of 3T scans with a reference 7T scan using the Statistical Parametric Mapping (SPM) tool box \cite{frackowiak2004human, friston1995spatial}\footnote{\url{https://www.fil.ion.ucl.ac.uk/spm/software/spm12/}}. Note that a bicubic interpolation is not required for this scenario as the 3T images are already registered to the 7T images and hence a mapping is learned from the registered 3T images to the 7T images. During the inference, the new 3T scan is registered to the reference 7T scan and is send through the learned network. This dataset addresses two issues 1.) a realistic image enhancement application where the a low quality MR image is enhanced to a high quality MR image, 2.) recently it has been argued that DW MRI images can be a substitute for Positron Emission Tomography (PET) images \cite{ohba2009diffusion, luboldt2008prostate, barchetti2016unenhanced} thereby confirming the versatility of our proposal.  
		
		\item 3T3T-T1: This scenario consists of the T1 Weighted MRI images of the same patients acquired by two different 3T scanners at different resolutions. The training and test strategy is similar to that of the 3T7T-DW. All the high resolution 3T scans are obtained from a customized Siemens 3T “Connectome Skyra”. Scans are acquired with a repetition time (TR) of 2400ms and an echo time (TE) of 2.14ms is used for acquisition. The dimension of the images in are $311\times 260\times 260$ in axial plane with a slice thickness of 0.7mm. The low-resolution 3T scans are acquired by Siemens Magnetom 3T scanner at a resolution of $136\times 113\times 113$ with a slice thickness of 1.6mm. The registration is performed similar to the procedure described for the above scenario. The LR 3T images are registered to the reference 3T HR image via the SPM tool box.   
	\end{enumerate}
	Quantitative results are reported in Table \ref{tab:real}. As can be observed, DNSP-EDSR-AP achieves superior performance over the state of the art EDSR network. Figures \ref{fig:3T7T} and \ref{fig:3T3T} show visual comparisons for example images from both the datasets. It can be observed that the DNSP-EDSR-AP enhanced image is closer to ground truth image compared to EDSR.
	
	\begin{table}[t]\captionsetup{labelfont={color=black},font={color=black}}
	\caption{\small{PSNR and SSIM comparisons for two real-world clinical datasets \cite{van2013wu}.}}
		\label{tab:real}
		\begin{center}
			\resizebox{.6\linewidth}{!}
			{\textcolor{black}{\begin{tabular}{cccc}
					\hline\hline
					\textbf{Method} &  \textbf{Dataset} &\textbf{PSNR} & \textbf{SSIM}\\\hline
					\multirow{2}{*}{EDSR} & 3T7T-DW & $24.91$ & $.8039$ \\
					& 3T3T-T1  & $26.35$ & $.8406$ \\
					\hline
					\multirow{2}{*}{DNSP-EDSR-AP} & 3T7T-DW & $\textbf{26.22}$ & $\textbf{.8581}$ \\
					& 3T3T-T1  & $\textbf{27.98}$ & $\textbf{.8670}$ \\
					\hline				
			\end{tabular}}}
		\end{center}
	\end{table}
	
	\subsection{Ablation Study Against a Total-Variation (TV) regularizer}
	\label{sec:TV}
\begin{figure}[t]
		\begin{center}
			\includegraphics[width=.8\linewidth]{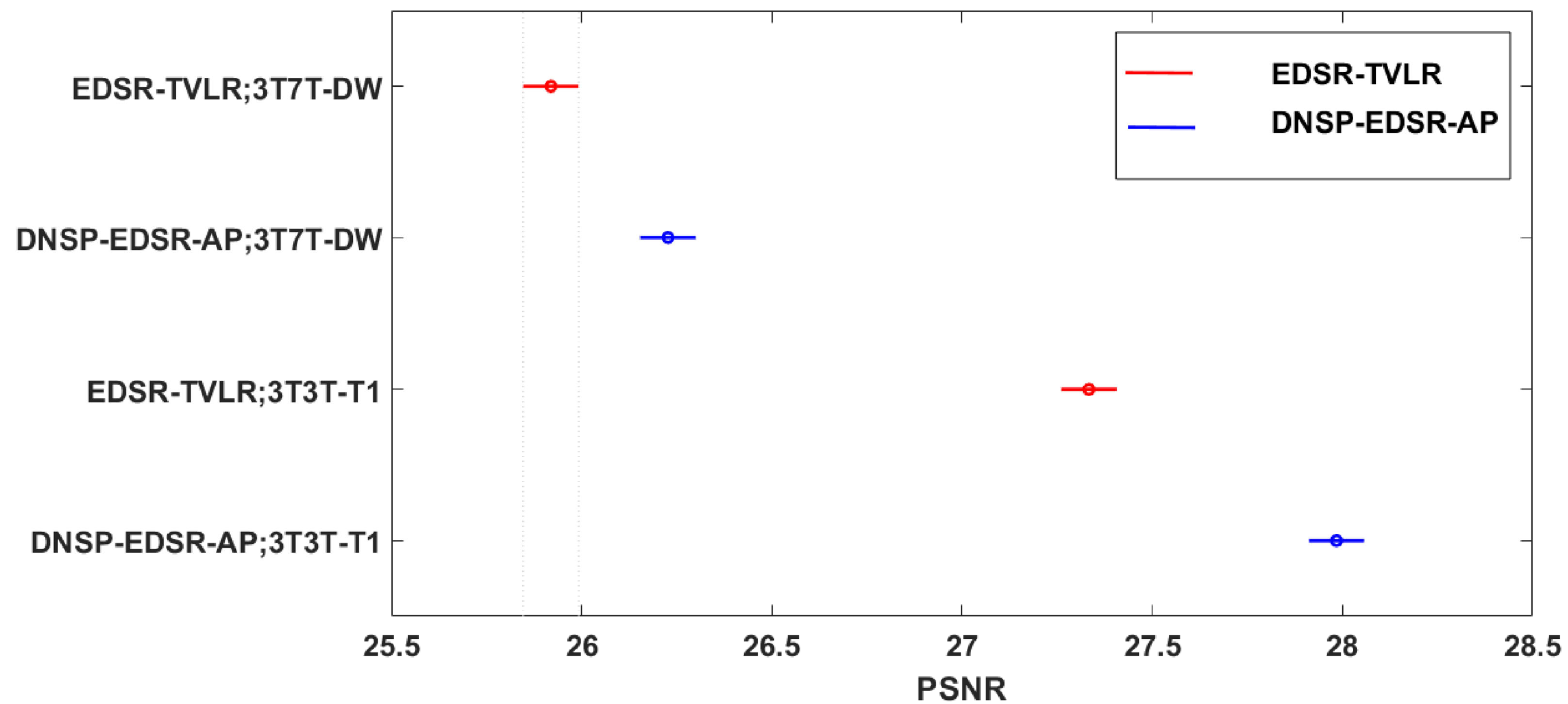}
		\end{center}\captionsetup{labelfont={color=black},font={color=black}}
		\caption{\small{2-way ANOVA comparing DNSP-EDSR-AP vs. EDSR-TVLR. The intervals represent the 95 $\%$ confidence intervals of PSNR values for a given configuration of method;dataset. Values reported for ANOVA across method factor are $df = 1$, $F = 143.37$, $p\ll .01$.}}
		\label{fig:ANOVA_real}
	\end{figure}

	Although a detailed ablation study is performed in Section \ref{sec:res}, the study is centered around the variants of our own method. To demonstrate the benefits of the proposed priors comprehensively, we perform an experiment that incorporates both low-rank and total-variation (TV) regularizers into a deep learning framework. We particularly chose TV and low-rank as this combination has proven successful for brain images in \cite{shi2015lrtv}. TV regularizer is used for recovering fine structures while suppressing noise. Our proposed sharpness enhancement measure does a similar job to a TV regularizer but more effectively as data-adaptive filters for a given dataset are learned where as a TV regularizer is generic and does not exploit any available training. Table \ref{tab:TV} reports the comparison of our method against the low-rank and TV priors incorporated with the EDSR network (called EDSR-TVLR) for the two real-world clinical datasets. Table \ref{tab:TV} reveals that DNSP-EDSR-AP performs the best.   
	Further, to confirm this statistically, Fig. \ref{fig:ANOVA_real} shows the 2-way ANOVA analysis for EDSR-TVLR and DNSP-EDSR-AP. It can be observed that DNSP-EDSR-AP is well separated from EDSR-TVLR for both the datasets.  
	\begin{table}[t]\captionsetup{labelfont={color=black},font={color=black}}
		\caption{\small{PSNR and SSIM comparisons against low-rank and TV regularizer for two real-world clinical datasets \cite{van2013wu}.}}
		\label{tab:TV}
		\begin{center}
			\resizebox{.6\linewidth}{!}
			{\textcolor{black}{\begin{tabular}{cccc}
					\hline\hline
					\textbf{Method} &  \textbf{Dataset} &\textbf{PSNR} & \textbf{SSIM}\\\hline
					\multirow{2}{*}{EDSR-TVLR} & 3T7T-DW & $25.92$ & $.8461$ \\
					& 3T3T-T1  & $27.34$ & $.8460$ \\
					\hline
					\multirow{2}{*}{DNSP-EDSR-AP} & 3T7T-DW & $\textbf{26.22}$ & $\textbf{.8581}$ \\
					& 3T3T-T1  & $\textbf{27.98}$ & $\textbf{.8670}$ \\
					\hline				
			\end{tabular}}}
		\end{center}
	\end{table}    
	
	\color{black}
	\section{Discussions and Conclusion}
	\label{sec:Conclusions}
	
	In this paper, we present a novel regularized deep network structure for MR image super-resolution, which excels in varying training regimes and experimental setups. This is accomplished by using two spatio-structural priors on the expected output HR image: 1) a low-rank prior, and 2) a sharpness prior. Our contributions include the development of new regularization terms that are inspired by the priors on the output of the network as well as tractable algorithmic methods to incorporate them in a deep learning set-up. We demonstrate the versatility of our method by experimental validation involving two widely used and highly competitive deep learning architectures for the SR problem. Because our priors are on the network output, the proposed DNSP method can be combined with many other deep SR networks as well.  
	
	Future work could develop and incorporate other meaningful priors such as those that are anatomically inspired \cite{oktay2018anatomically}. The interaction of prior induced regularization with specific network architectures can also be explored for speeding up network training and inference.

	\appendix
	
	\section{Back-propagation Derivations}
	First we derive the back-propagation equations for the loss function in Eq. (\ref{eq5})  which is given by:
	\label{sec:bpderive}
	\begin{equation}\label{eqobjective}
	l(\Theta) = \frac{1}{2}\|Y_{g} - F(X_{s}, \Theta)\|_{F}^{2} + \alpha R_{\delta}(Y) - \beta V(Y)
	\end{equation}
	where, $Y = F(X_{s}, \Theta)$, $\alpha$ and $\beta$ are positive regularization parameters. We learn $\Theta$ by minimizing $l(\Theta)$ using a stochastic gradient descent method \cite{werbos1994roots}. The weights ate each iteration are updated by the following rule
	\begin{equation}\label{eqUpdate1}
	\Theta^{t+1} = \Theta^{t} - \eta\frac{\partial l}{\partial\Theta^{t}}
	\end{equation}
	where, $t+1$ represents the iteration number, $\eta$ represents the learning rate for the stochastic gradient descent method and $\Theta^{t}$ represents the values of weights at previous iteration. As $\Theta = \{W_{k}^{l}, b_{k}^{l}\} \forall l,k$, following gradients are to be computed: $\frac{\partial E}{\partial w_{k}^{l}}$, $\frac{\partial E}{\partial b_{k}^{l}}$, where $w_{k}^{l}$ denotes an arbitrary scalar entry in filter $W_{k}^{l}$. For simplicity, let output image $Y$ be of dimension $N\times N$. The equation for computing the gradient of weight $w_{k}^{l}$ in any given layer $l$ is given by:
	\begin{equation}\label{eqW31}
	\frac{\partial E}{\partial w_{k}^{l}} = -(Y_{g} - Y)\diamond\frac{\partial Y}{\partial w_{k}^{l}} + \alpha D_{R_{\delta}}\diamond\frac{\partial Y}{\partial w_{k}^{l}} - \beta D_{V}\diamond\frac{\partial Y}{\partial w_{k}^{l}}\vspace{-2pt}
	\end{equation}
	where $D_{R_{\delta}} = -U\mbox{diag}\Bigg(-\frac{\sigma_{1}}{\delta^{2}}e^{-\sigma_{1}^{2}/2\delta^{2}},\ldots,-\frac{\sigma_{R}}{\delta^{2}}e^{-\sigma_{R}^{2}/2\delta^{2}}\Bigg)W^{T}$ is the gradient of $R_{\delta}(Y)$ and $D_{V}$ is the gradient for $V(Y)$. The complete expression for $D_{V}$ is given by:
	\begin{equation*}\label{eqDv1}
	D_{V} = [v_{i,j}], \mbox{  }v_{i,j} = d_{i,j} - \frac{1}{4}(d_{i-1,j} + d_{i+1,j} + d_{i, j-1} + d_{i, j+1})
	\end{equation*}
	\begin{equation}\label{eqDv2}\resizebox{0.85\linewidth}{!}{$d_{i,j} = \frac{2}{(N^{2})(N^{2} - 1)}\big(N^{2}p_{i,j} - \sum_{a}\sum_{b}p_{a,b} - \sum_{m}\sum_{n}(p_{m,n} - \frac{\sum_{m}\sum_{n}p_{m,n}}{N^{2}})\big)$}
	\end{equation}
	where $P = [p_{i,j}]$, and $P$ is obtained by convolving $Y$ with a $3\times 3$ laplacian operator $L$. Expression for $p_{i,j}$ is given by:
	\begin{equation}\label{eqDv31}\resizebox{0.88\linewidth}{!}{$
		p_{i,j} = y_{i,j} - \frac{1}{4}(y_{i-1,j} + y_{i+1,j} + y_{i, j-1} + y_{i, j+1}), \mbox{ where } Y = [y_{i,j}]$}
	\end{equation}
	Gradient for $R_{\delta}(Y)$ is derived in \cite{malek2014recovery}. Deriving expression for $D_{V}$ is mentioned below:\\
	To obtain laplacian of the output image $Y$, it is convolved with a $3\times 3$ filter $L = [[0 \mbox{ }- 1 \mbox{ }0]^{T} [-1 \mbox{   4 } \mbox{ }-1]^{T} [0\mbox{ } - 1\mbox{ } 0]^{T}]$. Let the laplacian be represented by $P$, where $p_{i,j}$ is given by Eq. (\ref{eqDv31}). Variance of laplacian $V(Y)$ is given by $V(Y) = var(P)$. Therefore,
	\begin{equation}\label{eqV}
	V(Y) = \frac{1}{N^{2} - 1}\sum_{i}\sum_{j}(p_{i,j} - \frac{\sum_{a}\sum_{b}p_{a,b}}{N^{2}})^{2}
	\end{equation}
	Now, gradient of $V_{Y}$ w.r.t $Y$ is obtained by following chain rule:
	\begin{align}\label{eqGv}
	v_{i,j} = \frac{\partial V}{\partial y_{i,j}} = &\frac{\partial V}{\partial p_{i,j}}.\frac{\partial p_{i,j}}{\partial y_{i,j}} + \frac{\partial V}{\partial p_{i,j-1}}.\frac{\partial p_{i,j-1}}{\partial y_{i,j}} + \frac{\partial V}{\partial p_{i,j+1}}.\frac{\partial p_{i,j+1}}{\partial y_{i,j}} \nonumber \\ &+ \frac{\partial V}{\partial p_{i-1,j}}.\frac{\partial p_{i-1,j}}{\partial y_{i,j}} + \frac{\partial V}{\partial p_{i + 1,j}}.\frac{\partial p_{i + 1,j}}{\partial y_{i,j}}
	\end{align}
	Note that $y_{i,j}$ can influence only $p_{i,j}$, $p_{i-1,j}$, $p_{i+1,j}$, $p_{i,j-1}$, $p_{i,j+1}$. Hence the chain rule is restricted only to these values as the partial derivative of all the $p_{i,j}$'s w.r.t $y_{i,j}$ is 0. It is straightforward to observe that
	\begin{equation}\label{eqPij}
	\frac{\partial p_{i,j}}{\partial y_{i,j}} = 1, \mbox{ }\frac{\partial p_{i,j-1}}{\partial y_{i,j}} = \frac{\partial p_{i,j+1}}{\partial y_{i,j}} = \frac{\partial p_{i+1,j}}{\partial y_{i,j}} = \frac{\partial p_{i-1,j}}{\partial y_{i,j}} = -1/4
	\end{equation}
	Substituting these values in Eq. (\ref{eqGv}) gives:
	\begin{equation}\label{eqGv1}
	v_{i,j} = \frac{\partial V}{\partial p_{i,j}} - \frac{1}{4}(\frac{\partial V}{\partial p_{i,j-1}} + \frac{\partial V}{\partial p_{i,j+1}} + \frac{\partial V}{\partial p_{i-1,j}}+ \frac{\partial V}{\partial p_{i + 1,j}})
	\end{equation}
	Now Eq. (\ref{eqDv2}) directly follows by taking derivative of $V(Y)$ in Eq. (\ref{eqV}) w.r.t $p_{i,j}$ to obtain $d_{i,j}$. \\
	\textbf{Back-Propagation Equations for Modified Loss Function:}
	The modified loss function is given by: 
	\begin{align}
	\label{eq:loss_mod1}
	E_{mod}(\Theta) = &\frac{1}{2}\|Y_{g} - F(X_{s}, \Theta)\|_{F}^{2} + \alpha R_{\delta}(Y) \nonumber \\ &- \beta V_{mod}(Y) + \gamma S(\Theta_{\mathcal{L}})
	\end{align}
	Following the lines of above derivation, the gradient of modified loss function w.r.t network parameter $w_{k}^{z}$ in layer $z$ is given by:  
	\begin{equation}\label{eq:bp_mod1}
	\resizebox{0.88\linewidth}{!}{$\frac{\partial E_{mod}}{\partial w_{k}^{z}} = -(Y_{g} - Y)\diamond\frac{\partial Y}{\partial w_{k}^{z}} + \alpha D_{R_{\delta}}\diamond\frac{\partial Y}{\partial w_{k}^{z}} - \beta D_{V_{mod}}\diamond\frac{\partial Y}{\partial w_{k}^{z}}$}
	\end{equation}  
	Note that network parameter $w_{k}^{z}$ does not depend on $S(\Theta_{\mathcal{L}})$, hence not reflected in back-propagation equations. The expression for $D_{R_{\delta}}$ remains same as described above. However, the expression for $D_{V_{mod}}$ differs from the fixed laplacian version. First, a set of $N_{\mathcal{L}}$ $3\times 3$ filters are used instead of a single $3\times 3$ laplacian filter wherein each filter is defined by set of coefficients $W_{\mathcal{L}_{k,m}}^{l}$, $(k,m) = \{-1,0,1\}$. Now, the expression for $D_{V_{mod}}$  is given by:
	\begin{equation}\label{eq:dvmod_1}
	D_{V_{mod}} = [v_{i,j}], \mbox{  }v_{i,j} = \frac{1}{N_{\mathcal{L}}}\sum_{l=1}^{N_{\mathcal{L}}}\sum_{k,m=-1}^{1}W_{\mathcal{L}_{k,m}}^{l} d_{i-k,j-m}^{l} ,
	\end{equation}
	\begin{equation*}\resizebox{\linewidth}{!}{$
		d_{i,j}^{l} = \frac{2}{(N^{2})(N^{2} - 1)}\big(N^{2}p_{i,j}^{l} - \sum_{a}\sum_{b}p_{a,b}^{l} - \sum_{m}\sum_{n}(p_{m,n}^{l} - \frac{\sum_{a}\sum_{b}p_{a,b}^{l}}{N^{2}})\big)$},
	\end{equation*}
	where $P^{l} = [p_{i,j}^{l}]$, and $P^{l}$ is obtained by convolving $Y$ with the $l^{th}$ $3\times 3$ learnable sharpness filter $W_{\mathcal{L}}^{l}$. Expression for $p_{i,j}^{l}$ is given by:
	\begin{equation*}
	p_{i,j}^{l} = \sum_{k,m=-1}^{1}W_{\mathcal{L}_{k,m}}^{l} y_{i,j}, \mbox{ and } Y = [y_{i,j}]
	\end{equation*} 
	The expression for $D_{V_{mod}}$ directly follows from the derivation of $D_{V}$ except for the fact that constant values in $D_{V}$ are replaced by the learnable filter parameters $W_{\mathcal{L}_{k,m}}^{l}$ and summed over all the $N_{\mathcal{L}}$ filters. \\
	The back-propagation equations for a given learnable sharpness filter parameter $W_{\mathcal{L}_{k,m}}^{l}$ is given by:
	\begin{equation}\label{eq:bp_llap1}
	\frac{\partial E_{mod}}{\partial W_{\mathcal{L}_{k,m}}^{l}} =  - \beta \frac{\partial V_{mod}(Y)}{\partial W_{\mathcal{L}_{k,m}}^{l}} + \gamma \frac{\partial S(\Theta_{\mathcal{L}})}{\partial W_{\mathcal{L}_{k,m}}^{l}}
	\end{equation}
	where $\frac{\partial S(\Theta_{\mathcal{L}})}{\partial W_{\mathcal{L}_{k,m}}^{l}}$ is given by:
	\begin{equation}
	\resizebox{.89\linewidth}{!}{$\frac{\partial S(\Theta_{\mathcal{L}})}{\partial W_{\mathcal{L}_{k,m}}^{l}} = 2 \sum_{a=1}^{N_{T}}(W_{\mathcal{L}}^{l}\circledast Sm_{a})\diamond Sm_{a}' - 2 \sum_{a=1}^{N_{T}}(W_{\mathcal{L}}^{l}\circledast Sh_{a})\diamond Sh_{a}'$}
	\end{equation}
	where $Sm_{a}' = [s_{a_{i-k,j-m}}]$, $s_{a_{i-k,j-m}}$ is the $(i-k,j-m)$ coefficient of $Sm$, $N_{T}$ is the number of training images. This expressions is derived by applying the chain rule - derivative of Frobenious norm followed by derivative of convolution operation w.r.t to the filter parameter. $Sh_{a}'$ is also defined similarly to $Sm_{a}'$. Following the same strategy, the expression for $\frac{\partial V_{mod}(Y)}{\partial W_{\mathcal{L}_{k,m}}^{l}}$ is given by:
	\begin{equation}
	\frac{\partial V_{mod}(Y)}{\partial W_{\mathcal{L}_{k,m}}^{l}} = D_{V_{mod}}\diamond Y'
	\end{equation}
	where $Y' = [y_{i-k,j-m}]$ and $D_{V_{mod}}$ is the same matrix as defined by Eq. (\ref{eq:dvmod_1}).
	
	%\vspace{-.4cm}
	%We present a novel regularized deep network structure for MR image superresolution, which excels in varying training regimes and experimental setups. This is accomplished by using two structural priors on the expected output HR image: 1) a low-rank prior, and 2) a sharpness prior. While we demonstrate improvements by employing SRCNN \cite{dong2016image} as our base network, our proposal is versatile and the proposed priors can be applied to extend other deep SR networks \cite{kim2016accurate, wang2015deep,timofte2017ntire,kim2016deeply} as well.
	
	\bibliographystyle{IEEEtran}
	\bibliography{refs}
\end{document}